\documentclass[runningheads]{llncs}
\usepackage[T1]{fontenc}
\usepackage{mathtools}
\usepackage{amsmath}
\usepackage{graphicx}
\usepackage[hyphens]{url} 
\usepackage[colorlinks]{hyperref} 
\hypersetup{breaklinks=true, pdfencoding=auto}

\usepackage{cleveref}
\usepackage{subcaption}
\usepackage{todonotes}
\usepackage{orcidlink}
\usepackage{enumitem}
\usepackage{graphicx}
\usepackage{amssymb}
\usepackage{algorithmicx}
\usepackage[noend]{algpseudocode}
\usepackage[percent]{overpic} 
\usepackage{algorithm}
\usepackage[export]{adjustbox}
\usepackage{bbding}

\usepackage{tcolorbox}
\tcbuselibrary{skins}
\newtcolorbox{rqbox}[1]{enhanced,attach boxed title to top center={yshift=-2.2mm},fonttitle=\sffamily\bfseries,coltitle=black,boxed title style={colframe=white,colback=white,top=-0.2em,bottom=-0.4em},colback=white,sharp corners,boxrule=0.3mm,title={#1},left=2mm,right=2mm,bottom=2mm,enlarge top by=-0.5em}

\usepackage{tikz}
\usetikzlibrary{shapes, fit}
\usepackage{color}

\begin{document}
\title{Analysis of input-output mappings in coinjoin transactions with arbitrary values}
\titlerunning{Analysis of input-output mappings in coinjoins}
%
\author{Jiri Gavenda\orcidlink{0009-0009-9026-781X}\Envelope \and
Petr Svenda\orcidlink{0000-0002-9784-7624} \and Stanislav Bobon \and Vladimir Sedlacek\orcidlink{0000-0003-2409-661X}
}
\authorrunning{Gavenda et al.}
%
\institute{Masaryk University, Czechia \\
\email{\{gavenda,xsvenda,xbobon,vlada.sedlacek\}@mail.muni.cz}}

\maketitle              
\vspace{-0.45cm}
\begin{abstract}

A coinjoin protocol aims to increase transactional privacy for Bitcoin and Bitcoin-like blockchains via collaborative transactions, by violating assumptions behind common analysis heuristics. Estimating the resulting privacy gain is a crucial yet unsolved problem due to a range of influencing factors and large computational complexity. 

We adapt the BlockSci on-chain analysis software to coinjoin transactions, demonstrating a significant (10-50\%) average post-mix anonymity set size decrease
for all three major designs with a central coordinator: Whirlpool, Wasabi 1.x, and Wasabi 2.x. The decrease is highest during the first day and negligible after one year from a coinjoin creation. 

Moreover, we design a precise, parallelizable privacy estimation method, which takes into account coinjoin fees, implementation-specific limitations and users' post-mix behavior.
We evaluate our method in detail on a set of emulated and real-world Wasabi 2.x coinjoins and extrapolate to its largest real-world coinjoins with hundreds of inputs and outputs. We conclude that despite the users' undesirable post-mix behavior, correctly attributing the coins to their owners is still very difficult, even with our improved analysis algorithm.

\end{abstract}

\section{Introduction}
A coinjoin is a collaborative cryptocurrency transaction attempting to improve the users' privacy.
A good estimation of the resulting level of anonymity is crucial for multiple parties. The users need a metric to decide when their funds are sufficiently mixed to stop participating, as every iteration incurs a non-zero cost paid in mining and coordination fees. Meanwhile, chain analysts -- such as law enforcement agencies (LEAs) -- need data to interpret digital evidence in case of tracking stolen funds or funds coming from a criminal activity.
In both of these scenarios, anonymity is typically computed based on a suite of heuristics with no proper data-supported evidence, partially due to the inherently missing ground truth for real coinjoin transactions. 

One option to estimate anonymity is to enumerate all possible input-output mappings, as done in \cite{maurer2017anonymous} and \cite{boltzmann}. However, all currently published approaches suffer from potentially false assumptions (e.g., number of inputs coming from one wallet) and imprecisions (e.g., oversimplification of fees). We believe this methodology needs to be adapted to the design and implementation of the specific coinjoin in question.
  
Additionally, the privacy depends not only on the transaction on its own, but also on the subsequent on-chain behavior of all participating users. Even if a given output initially ``hides'' within a set of outputs of the same denomination, subsequent transactions often reveal its owner -- due to consolidations (joining multiple outputs into one) or linking with real-world identities (e.g., sending an output to a regulated exchange).

\vspace{0.5cm} 
\noindent \textbf{Research questions.} 
To investigate the privacy properties of real coinjoins, we ask the following research questions:
\begin{enumerate}[leftmargin=3.1em]
\itemsep0em
\item [\textbf{RQ1:}] What is the post-mix behavior of users participating in Wasabi 1.x, 2.x and Whirlpool coinjoin designs, and how does it affect the initially obtained anonymity of mixed outputs?

\item [\textbf{RQ2:}] How to algorithmically evaluate the anonymity obtained from coinjoin participation when additional information about the users' on-chain behavior and client implementation is available? 
\end{enumerate}

\noindent \textbf{Contributions.} To answer our research questions, we contribute the following:
\begin{itemize}
    \itemsep0em
    \item We extend the major open-source BlockSci project \cite{blocksci20} for detection and analysis of coinjoins with extraction of behavioral consolidation patterns of real on-chain users, highlighting eventual stabilization of the anonymity set loss over time (\Cref{sec:blocksci_coinjoinloss}). 
    \item We design a precise generic method to compute all possible mappings between sets of coinjoin inputs and outputs. It incorporates hard limitations of client software implementation, participation fees, and consolidation behavior observed for on-chain users (\Cref{sec:mappings_enum}). 
    \item We evaluate the method on real coinjoins, showing privacy gains with practical implications both for the users as well as forensic analysts interpreting past transactions (\Cref{sect:ww2_mapping_enums}). 
\end{itemize}

The extended BlockSci project is available at \url{https://github.com/crocs-muni/blocksci}. Data processing scripts and all results are available under a permissive license at \url{https://github.com/crocs-muni/coinjoin-mappings}.

\section{Coinjoin transactions and mappings}
\label{sec:background}

A \emph{coinjoin transaction} (or just a \emph{coinjoin}) \cite{maxwell_cj} is created collaboratively by multiple users with the goal of gaining privacy by obfuscating the link between the addresses of their inputs and outputs. In a typical coinjoin, each user owns multiple inputs and multiple outputs. Repeating values and multiple possibilities of decomposing the input sum then create uncertainty about which inputs and outputs are linked together. \href{subfig:coinjoin}{Figure 1a} shows a simple example (omitting fees).

\subsection{Our coinjoin model}\label{sec:model}
We use a notation partially compatible with \cite{maurer2017anonymous} but modify it to consider fees. 
We denote the set of input and output coins by $I$ and $O$, respectively, and fix a mapping $v: I \cup O \rightarrow \mathbb{N}$ that assigns numerical values\footnote{These are written as fractions of BTC, but correspond to integer amounts of satoshis.} to coins\footnote{By abuse of language, we sometimes do not distinguish between a coin and its value.}. A \emph{coinjoin} is then a triple $T = (I, O, v)$.
For a user $u \in U$, we let $\epsilon_u \geq 0$, $\chi_u$ be their mining and coordination\footnote{This fee could be negative if the user receives money from others. In particular, there can be a special user $c$ called a \emph{coordinator} with $\epsilon_c = 0$ and $\chi_c = - \sum_{u \in U \setminus\{c\}} \chi_u$.} fees, respectively.

\begin{definition}
 For $u\in U$, $I_u \subseteq I$, $O_u \subseteq O$, we call $(I_u, O_u)$ a \emph{sub-mapping} if 
\begin{equation}\label{eq:coinjoin-def}
    \sum_{i \in I_u}v(i) = \sum_{o \in O_u}v(o) + \epsilon_u + \chi_u.
\end{equation}
\end{definition}

\begin{definition}
 A \emph{mapping} is a set of sub-mappings $\{(I_1,O_1),\dots,(I_k,O_k)\}$ such that for each $u,u' \in U$, $u\neq u'$, we have $I_u \subseteq I$, $O_u \subseteq O$, $I_u \cap I_{u'} = \emptyset$, $O_u \cap O_{u'} = \emptyset$.
\end{definition}


In this paper, we study the different mappings admissible for a given coinjoin~$T$.
We denote the set of all such mappings by $\mathcal{M}_T$, or just $\mathcal{M}$ when $T$ is clear from the context. Note that while $T$ is public information, $\mathcal{M}_T$ is typically difficult to compute, and the number of users is not even constant across the mappings. \href{subfig:mappings}{Figure 1b} shows 4 out of the 24 mappings for the example coinjoin. 
\begin{figure}
    \centering
    \begin{subfigure}{0.4\textwidth}
    \label{subfig:coinjoin}
    \begin{tikzpicture}
        \node[draw, fill=gray!20, minimum size=0.6cm] (I1) at (0,2.1) {8};
        \node[draw, fill=gray!20, minimum size=0.6cm] (I2) at (0,1.4) {6};
        \node[draw, fill=gray!20, minimum size=0.6cm] (I3) at (0,0.7) {3};
        \node[draw, fill=gray!20, minimum size=0.6cm] (I4) at (0,0) {3};
        
        \node[draw, fill=gray!20, minimum size=0.6cm] (O1) at (3,2.45) {6};
        \node[draw, fill=gray!20, minimum size=0.6cm] (O2) at (3,1.75) {6};
        \node[draw, fill=gray!20, minimum size=0.6cm] (O3) at (3,1.05) {4};
        \node[draw, fill=gray!20, minimum size=0.6cm] (O4) at (3,0.35) {2};
        \node[draw, fill=gray!20, minimum size=0.6cm] (O5) at (3,-0.35) {2};
        
        \node at (0, 2.8) {\textbf{Inputs}};
        \node at (3, 3.2) {\textbf{Outputs}};

        \node[draw=none] (M) at (1.5,1.05) {};
        
        \draw (I1.east) to[out=0, in=180] (M.center);
        \draw (I2.east) to[out=0, in=180] (M.center);
        \draw (I3.east) to[out=0, in=180] (M.center);
        \draw (I4.east) to[out=0, in=180] (M.center);
        
        \draw (M.center) to[out=0, in=180] (O1.west);
        \draw (M.center) to[out=0, in=180] (O2.west);
        \draw (M.center) to[out=0, in=180] (O3.west);
        \draw (M.center) to[out=0, in=180] (O4.west);
        \draw (M.center) to[out=0, in=180] (O5.west);
        
    \end{tikzpicture}
    \caption{A coinjoin transaction.}
    \end{subfigure}
    \hfill
    \begin{subfigure}{0.5\textwidth}
    \label{subfig:mappings}
    \begin{tikzpicture}
        \node[draw, fill=blue!20, minimum size=0.3cm] (I1) at (0,1.2) {\tiny 8};
        \node[draw, fill=red!20, minimum size=0.3cm] (I2) at (0,0.8) {\tiny 6};
        \node[draw, fill=green!20, minimum size=0.3cm] (I3) at (0,0.4) {\tiny 3};
        \node[draw, fill=green!20, minimum size=0.3cm] (I4) at (0,0) {\tiny 3};
        
        \node[draw, fill=red!20, minimum size=0.3cm] (O1) at (2,1.4) {\tiny 6};
        \node[draw, fill=green!20, minimum size=0.3cm] (O2) at (2,1) {\tiny 6};
        \node[draw, fill=blue!20, minimum size=0.3cm] (O3) at (2,0.6) {\tiny 4};
        \node[draw, fill=blue!20, minimum size=0.3cm] (O4) at (2,0.2) {\tiny 2};
        \node[draw, fill=blue!20, minimum size=0.3cm] (O5) at (2,-0.2) {\tiny 2};

        \node[draw=none] (M) at (1,0.6) {};
        
        \draw (I1.east) to[out=0, in=180] (M.center);
        \draw (I2.east) to[out=0, in=180] (M.center);
        \draw (I3.east) to[out=0, in=180] (M.center);
        \draw (I4.east) to[out=0, in=180] (M.center);
        
        \draw (M.center) to[out=0, in=180] (O1.west);
        \draw (M.center) to[out=0, in=180] (O2.west);
        \draw (M.center) to[out=0, in=180] (O3.west);
        \draw (M.center) to[out=0, in=180] (O4.west);
        \draw (M.center) to[out=0, in=180] (O5.west);
        
    \end{tikzpicture}
    \vspace{0.1cm}
    \begin{tikzpicture}
        \node[draw, fill=blue!20, minimum size=0.3cm] (I1) at (0,1.2) {\tiny 8};
        \node[draw, fill=red!20, minimum size=0.3cm] (I2) at (0,0.8) {\tiny 6};
        \node[draw, fill=green!20, minimum size=0.3cm] (I3) at (0,0.4) {\tiny 3};
        \node[draw, fill=green!20, minimum size=0.3cm] (I4) at (0,0) {\tiny 3};
        
        \node[draw, fill=blue!20, minimum size=0.3cm] (O1) at (2,1.4) {\tiny 6};
        \node[draw, fill=red!20, minimum size=0.3cm] (O2) at (2,1) {\tiny 6};
        \node[draw, fill=green!20, minimum size=0.3cm] (O3) at (2,0.6) {\tiny 4};
        \node[draw, fill=blue!20, minimum size=0.3cm] (O4) at (2,0.2) {\tiny 2};
        \node[draw, fill=green!20, minimum size=0.3cm] (O5) at (2,-0.2) {\tiny 2};

        \node[draw=none] (M) at (1,0.6) {};
        
        \draw (I1.east) to[out=0, in=180] (M.center);
        \draw (I2.east) to[out=0, in=180] (M.center);
        \draw (I3.east) to[out=0, in=180] (M.center);
        \draw (I4.east) to[out=0, in=180] (M.center);
        
        \draw (M.center) to[out=0, in=180] (O1.west);
        \draw (M.center) to[out=0, in=180] (O2.west);
        \draw (M.center) to[out=0, in=180] (O3.west);
        \draw (M.center) to[out=0, in=180] (O4.west);
        \draw (M.center) to[out=0, in=180] (O5.west);
        
    \end{tikzpicture}
    \hfill
    \begin{tikzpicture}
        \node[draw, fill=blue!20, minimum size=0.3cm] (I1) at (0,1.2) {\tiny 8};
        \node[draw, fill=green!20, minimum size=0.3cm] (I2) at (0,0.8) {\tiny 6};
        \node[draw, fill=blue!20, minimum size=0.3cm] (I3) at (0,0.4) {\tiny 3};
        \node[draw, fill=blue!20, minimum size=0.3cm] (I4) at (0,0) {\tiny 3};
        
        \node[draw, fill=blue!20, minimum size=0.3cm] (O1) at (2,1.4) {\tiny 6};
        \node[draw, fill=blue!20, minimum size=0.3cm] (O2) at (2,1) {\tiny 6};
        \node[draw, fill=green!20, minimum size=0.3cm] (O3) at (2,0.6) {\tiny 4};
        \node[draw, fill=green!20, minimum size=0.3cm] (O4) at (2,0.2) {\tiny 2};
        \node[draw, fill=blue!20, minimum size=0.3cm] (O5) at (2,-0.2) {\tiny 2};

        \node[draw=none] (M) at (1,0.6) {};
        
        \draw (I1.east) to[out=0, in=180] (M.center);
        \draw (I2.east) to[out=0, in=180] (M.center);
        \draw (I3.east) to[out=0, in=180] (M.center);
        \draw (I4.east) to[out=0, in=180] (M.center);
        
        \draw (M.center) to[out=0, in=180] (O1.west);
        \draw (M.center) to[out=0, in=180] (O2.west);
        \draw (M.center) to[out=0, in=180] (O3.west);
        \draw (M.center) to[out=0, in=180] (O4.west);
        \draw (M.center) to[out=0, in=180] (O5.west);
        
    \end{tikzpicture}
    \hfill
    \begin{tikzpicture}
        \node[draw, fill=blue!20, minimum size=0.3cm] (I1) at (0,1.2) {\tiny 8};
        \node[draw, fill=green!20, minimum size=0.3cm] (I2) at (0,0.8) {\tiny 6};
        \node[draw, fill=green!20, minimum size=0.3cm] (I3) at (0,0.4) {\tiny 3};
        \node[draw, fill=green!20, minimum size=0.3cm] (I4) at (0,0) {\tiny 3};
        
        \node[draw, fill=green!20, minimum size=0.3cm] (O1) at (2,1.4) {\tiny 6};
        \node[draw, fill=blue!20, minimum size=0.3cm] (O2) at (2,1) {\tiny 6};
        \node[draw, fill=green!20, minimum size=0.3cm] (O3) at (2,0.6) {\tiny 4};
        \node[draw, fill=green!20, minimum size=0.3cm] (O4) at (2,0.2) {\tiny 2};
        \node[draw, fill=blue!20, minimum size=0.3cm] (O5) at (2,-0.2) {\tiny 2};

        \node[draw=none] (M) at (1,0.6) {};
        
        \draw (I1.east) to[out=0, in=180] (M.center);
        \draw (I2.east) to[out=0, in=180] (M.center);
        \draw (I3.east) to[out=0, in=180] (M.center);
        \draw (I4.east) to[out=0, in=180] (M.center);
        
        \draw (M.center) to[out=0, in=180] (O1.west);
        \draw (M.center) to[out=0, in=180] (O2.west);
        \draw (M.center) to[out=0, in=180] (O3.west);
        \draw (M.center) to[out=0, in=180] (O4.west);
        \draw (M.center) to[out=0, in=180] (O5.west);
        
    \end{tikzpicture}
    \caption{Input-output mappings.}
    \end{subfigure}
    
    \caption{An example of a coinjoin with arbitrary output denominations, without fees and with 4 of 24 possible input-output mappings visualized.}
    \label{fig:enter-label}
\end{figure}
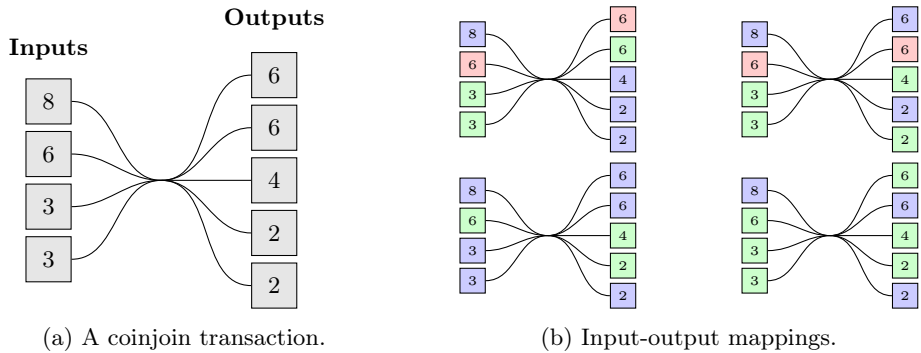

\subsection{Coinjoin implementations}
On a very high level, a coinjoin generally consists of several phases:
\begin{itemize}
    \item Input selection: each user registers coins that they would like to mix (typically done by the wallet). 
    \item Output selection: each wallet determines
    how to decompose the sum of user's inputs into outputs of feasible denominations (either fixed or arbitrary). 
    \item Signing: each user confirms that they agree with the result.
\end{itemize}

A typical example of a coinjoin implementation with fixed values is Samourai Whirlpool~\cite{sw} based on the ZeroLink protocol \cite{zerolink}. Coinjoins created with Whirlpool have the same number (typically 5 or 8) of inputs and outputs of the same denomination (0.5, 0.05, 0.01, or 0.001 BTC), except for some of the inputs having a slightly higher value to pay the mining fees. 

Wasabi 1.0 \cite{ww1} is also based on Zerolink. Its inputs can have any value, but each user has a single output of a fixed size and possibly a second output for a leftover value (``change''). From version 1.1, each transaction can have multiple mixed denominations, and each user can have one output per denomination.

Wasabi 2.x \cite{ww2} is based on the WabiSabi protocol \cite{ficsor2021wabisabi}, allowing coinjoins with arbitrary values. Users can register any values as inputs and outputs, but their wallets compute possible output denominations using a predefined algorithm.  Limiting the number of output denominations assures a high probability of multiple users registering the same output values. 

Another example of an implementation with arbitrary values is JoinMarket~\cite{jm}. While the previous implementations have a coordinator facilitating the creation of the transaction, JoinMarket instead features users of two kinds -- makers and takers. Takers are the ones who want to gain anonymity by mixing, while makers provide their coins for mixing for a fee. A taker selects multiple makers, then creates the transaction and asks the makers to sign it. Thus, the taker knows the mapping of the transaction, while the makers do not. 

\subsubsection{Fee structures}

These implementations differ in their fee structure. The fees are not only economically relevant to the users, but also impact the reconstruction of mappings, as a user's output values may not always match their input values.

Typically, there are two types of fees: 
\begin{enumerate}
    \item Mining fees paid to miners of the block in which the transaction is included.
    \item Coordination fees paid to the transaction coordinator or other users.
\end{enumerate} 

In Samourai Whirlpool, the coordination fee is paid in a special transaction TX0 preceding the actual mixing. TX0 serves to split the original values into the standard values used in Whirlpool. The coordination fee is a percentage of the standard denomination, typically around 5\% of the value, but tens of such outputs can be obtained from a single TX0. The outputs of TX0 are slightly higher than the standard value to pay for mining fees. To allow repeated mixing, mining fees are paid in full by the inputs coming directly from TX0, while the inputs coming from previous coinjoins do not incur any fees. 

Both Wasabi 1.x and Wasabi 2.x split the cost of mining fees fairly between the users based on the number of inputs and outputs they register. In Wasabi 1.x, the coordination fee depends on the size of the standard outputs and the number of outputs with the same value (typically about 0.15\% of the standard output value). In Wasabi 2.x, until version 2.2.0, a coordinator could charge a fee as a percentage of the input value. The default coordinator charged 0.3\%, but let inputs of less than 0.01 BTC and inputs coming from a previous coinjoin mix for free.
All Wasabi 2.x versions have to deal with decomposition leftover: since a precise decomposition of the input sum into a set of output values is not always possible, and due to mining fees, it may not be economically reasonable for the user to get the leftover value (``change''). The coordinator can claim this change if enough users leave them in a given round; otherwise, it is left for miners.

A JoinMarket's taker pays a fee to the makers as a reward for letting him use their funds. The fee currently averages at around 0.001\% of the value provided by a maker. The taker typically also pays the mining fee for the whole transaction.

\subsection{Previous work}

A major privacy concern of Bitcoin is the possibility of clustering addresses based on their ownership. Finding the owner of one of the addresses reveals the owner of the whole cluster. Many papers were dedicated to simple yet efficient heuristics for such address clustering and on-chain activity analysis \cite{lnk_androulaki,blocksci20,lnk_maesa2016uncovering,meiklejohn2013fistful,lnk_Reid2013,lnk_dorit}. 

As the adoption of coinjoins started growing, so did the need to adapt the on-chain clustering heuristics. Most attempts in this area focused simply on the detection and avoidance of coinjoins and gathering statistical data about these transactions \cite{dumplings,joinmarket2017,schnoering2023heuristics,coinjoin_adoption_22,detectWasabi22,wahrstatter2024reducing}. We show how the on-chain heuristics can be used to facilitate the analysis of the post-mix behavior of coinjoin users. 

Goldfeder et al. \cite{DBLP:journals/popets/GoldfederKRN18} analyzed the possibility of linking addresses to user identities using cookies on merchants' websites. They further proposed an intersection attack allowing the linking of inputs and outputs of coinjoins. However, this attack assumes a perfect clustering of the transaction inputs. 

Two works \cite{boltzmann,maurer2017anonymous} have analyzed input-output ownership in coinjoins by looking at the possible mappings. In both of these works, the entropy and probability computation works only under the assumption that all the mappings have the same likelihood of occurrence. This assumption does not hold in practice, as some mappings never occur, and some are more likely than others due to the implementations' limits and users' behavior.

Maurer et al. \cite{maurer2017anonymous} additionally analyzed the upper bound of the complexity of the enumeration of all coinjoin mappings in a case without fees. They describe how such an enumeration can be used to compute the Shannon entropy of the transaction and the probability that the same user owns an input-output pair. 

The same entropy and input-output link probability are also computed by the Boltzmann tool \cite{boltzmann}, which can work with real transactions with mining fees as well as with the fee structure of JoinMarket. It does not work explicitly with the individual mappings, but speeds up the computation by computing only a matrix of the link probabilities. 

We follow up and show how to enumerate mappings with a more precise fee computation and considering properties of real coinjoins, without assuming that all mappings have equal probability. 

\subsection{Attacker model}
We assume an observer performing analysis of all the information recorded on the public blockchain, as well as private information obtained by actively participating in transactions and interacting with public APIs of the discussed coinjoin implementations. A typical example is a chain analysis company providing anti-money laundering checks for regulated financial institutions or support for law enforcement investigations.  

\subsection{Anonymity}
Notions of anonymity can differ and choosing the ``correct'' one can lead to fundamental philosophical questions. We use a pragmatic definition: The anonymity of a coinjoin output is the entropy (amount of uncertainty) that an attacker has regarding correctly attributing this output to its true owner. Clearly, this depends on the attacker's knowledge; when not specified otherwise, we will assume they only have public knowledge.

\section{Analysis of post-mix anonymity set loss}
\label{sec:blocksci_coinjoinloss}

A passive attacker (analyst) can retrospectively obtain all past coinjoins recorded on the blockchain. If analyzed in separation, the basic source of uncertainty is the \emph{anonymity set} of an output $o$, defined as $A(o) \coloneqq \{o' \in O \mid v(o') = v(o)\}$.
In this section, we extend the on-chain analysis tool BlockSci \cite{blocksci20} to detect post-mix merges (consolidations) of coinjoin outputs using common ownership heuristics. Then we use the consolidations to quantify the decrease in the effective anonymity set size of each coinjoin, which impacts privacy not only for the owner of the consolidated outputs, but also for all other users. 
The detected consolidations serve as an input to our mappings enumeration algorithm (\Cref{sect:ww2_mapping_enums}).

\subsubsection{BlockSci project and its limitations}

BlockSci \cite{blocksci20} is an open-source project by researchers at Princeton University that allows on-chain analysis of Bitcoin and Bitcoin-like blockchains. It provides efficient and easy-to-use programmatic access to all blockchain data without being tailored to a specific use, using two interfaces: a Python API and a C++ API. The Python API is designed for simpler and intuitive exploration using a Jupyter Notebook, while the C++ API is designed for performance-heavy and highly parallelizable tasks. The parallelization utilizes the \emph{MapReduce} framework~\cite{mapreduce} and always-in-RAM placement of crucial data structures to perform fast and large-scale analysis over a direct acyclic graph (DAG) and related transaction metadata extracted from on-chain as well as off-chain sources. The \emph{Map} function is parallelized on input data segments and performs detection of coinjoins or address clustering using multiple CPUs. The \emph{Reduce} function combines the \emph{Map}'s outputs, e.g., collating detected clusters of addresses.

The original version of BlockSci suffers from three main limitations:
\begin{itemize}
    \item it does not analyze the context of coinjoin-related transactions,
    \item its address clusterer is insufficient for coinjoin analysis,
    \item it is no longer actively maintained (since 2020) and uses outdated heuristic algorithms, limiting its utility.
\end{itemize}
We address all these limitations and present an updated BlockSci project with added coinjoin detection algorithms, coinjoin context analyses, and deployment scripts at \url{https://github.com/crocs-muni/blocksci}\footnote{The original upstream repository is not accepting any new pull requests.}.

\subsection{An improved coinjoin detection algorithm}
The original BlockSci already contained a simplistic coinjoin detection algorithm used to \emph{avoid} processing the coinjoins in subsequent address clustering analyses. While the algorithm might have been sufficient in 2019, it fails to properly classify many coinjoins created by later designs, resulting in a massive number of both false positives and false negatives.    

We extended BlockSci with three algorithms for improved coinjoin detection based on the work of Stütz et al.~\cite{coinjoin_adoption_22}. For Wasabi 2.x, we applied two more rules:
\begin{itemize}
    \item the inputs and outputs must belong to at least five different addresses,
    \item for transactions executed after May 1st, 2024, there are at least 20 inputs (unlike 50 before) -- to account for the emergence of new, smaller coordinators after the official Wasabi 2.x coordinator (zkSNACKs) was stopped. 
\end{itemize}

The decreased threshold for a number of inputs naturally resulted in an increased number of false positives, which we mitigated by detecting candidates with a missing connection to any other coinjoin and/or exhibiting a level of address reuse higher than 70\%. We manually verified the candidate false positives and placed them on the continuously maintained exclusion list.  

The final detection accuracy was verified against the Dumplings project \cite{dumplings} as well as a list of all finished coinjoins obtained by actively querying running coinjoin coordinators. We obtained a perfect match for Whirlpool transactions and a near-perfect match for Wasabi 1.x and Wasabi 2.x transactions, with differences identified manually as misclassifications by the Dumplings project.

\subsection{Anonymity degradation over time}
\label{sect:anon_degradation_alg}

The initial anonymity of an output obtained at the time of participation in a coinjoin is not permanent. Depending on the actions carried out by the owner as well as other users, it will likely degrade over time. For example, if a user consolidates two (or more) outputs from one coinjoin, the \emph{common input ownership} heuristic links both outputs together into the same address cluster, decreasing their anonymity. 
Crucially, this also decreases the anonymity of all outputs with the same denomination, as the number of indistinguishable outputs decreases.

A significant obstacle to determining the exact anonymity of outputs in real coinjoins is the lack of knowledge about the post-mix behavior of each separate user, which is exactly the information the user wants to hide by using a coinjoin. However, we can still study the statistical patterns of users' behavior by using post-mix consolidation heuristics. 

For a coinjoin output $o$, we define its anonymity set loss after $d$ days 
as $$A_d(o)\coloneqq \frac{|\{o\in A(o) \mid o \text{ has been consolidated within $d$ days after the coinjoin \}}|}{|A(o)|},$$
and extend this to the whole transaction by averaging over all outputs\footnote{This is the same as a weighted average over all denominations.}: 
$$A_d \coloneqq \frac{1}{|O|} \sum_{o\in O} A_d(o).$$
By a consolidation, we mean a non-coinjoin transaction that includes at least two inputs incoming from (not necessarily distinct) coinjoin transactions. (We disregard transactions with only a single coinjoin input, as these are not linking mixed outputs of the same user).

The initial $A_0 = 0$ represents anonymity set loss when no future user actions are considered yet, while $A_\infty$ considers all consolidation transactions to present\footnote{As of 6 April 2025.}. As we show on real coinjoin data in \Cref{sect:degradation_results}, the consolidation patterns vary significantly in time and among different coinjoin designs. 

\begin{algorithm}
    \centering
    \begin{algorithmic}
        \Function{get\_consolidated\_outputs}{$cj, t, CJ$}
            \Comment{$cj,t$: txs, $CJ$: all coinjoins}
            \If{$|$\Call{outputs}{$CJ$} $\cap$ \Call{inputs}{$t$}$|$ $\geq2$}
                \State \Return  \Call{outputs}{$cj$} $\cap$ \Call{inputs}{$t$}
                \Comment{All consolidated $cj$'s outputs}
            \Else
                \State \Return $\emptyset$ 
                \Comment{No consolidation detected}
            \EndIf    
        \EndFunction

\vspace{0.2cm}
    
        \Function{compute\_loss}{$d$, $CJ$} \Comment{$d$: days interval, $CJ$: all coinjoins}
            \State $H \gets \{\}$ \Comment{Initialize dictionary of losses}
            \ForAll{$cj \in CJ$}
                \State $X \gets \emptyset$ \Comment{Initialize set of detected consolidations}
                \ForAll{$t \in$ \Call{get\_txs\_in\_timeframe}{$time(cj)$, $d$}}
                    \State $O \gets $ \Call{get\_consolidated\_outputs}{$cj, t, CJ$} 
                    \State $X.\Call{append}{O}$
                \EndFor
                \State $H[cj] \gets |X| / |$\Call{outputs}{$cj$}$|$
                \Comment{Ratio of consolidated to all $cj$'s outputs}
            \EndFor
            \State \Return $H$  
        \EndFunction
    \end{algorithmic}
    \caption{Anonymity set loss due to coinjoin outputs consolidation.} 
    \label{fig:entropy_loss_alg}
    
\end{algorithm}

Our \Cref{fig:entropy_loss_alg} first computes the dictionary by linking specific output denominations to the number of their occurrences for each coinjoin. Then, we search for any transaction (up to $d$ days after the initial coinjoin for computing $A_d$) that has $\geq 2$ inputs from any of the detected coinjoins and lower the corresponding number of occurrences in the dictionary. Outputs mixed again by a future coinjoin are automatically excluded from consolidations as assumptions for common input heuristics are trivially invalidated. We optimize our analysis by avoiding repeated iteration over huge DAG structures and parallelize it using the \emph{MapReduce} paradigm. 

\Cref{fig:entropy_loss_alg} is \emph{conservative} and will typically \emph{overestimate} $A_d$ due to two factors. Firstly, we assume that any consolidation transaction provides an actionable anonymity set loss -- i.e., an analyst can attribute the merged outputs to a specific user from a set of known coinjoin inputs owners. In practice, the analyst will be able to do so only for some fraction of all consolidation transactions, resulting in a smaller anonymity set loss than computed.    
Secondly, the inclusion of any misclassified consolidation transaction (e.g., structurally different coinjoin designs) will also artificially increase the computed $A_d$, despite not providing truly actionable information to the analyst. However, the occasional inclusions of coinjoins with a small number of outputs only have a small impact on the aggregate statistics for the anonymity set loss presented in \Cref{sect:degradation_results}. If the number of these false consolidations starts to rise in the future (e.g., due to a new coinjoin design with a large number of inputs), the coinjoin detection and consolidation detection algorithms will need to be updated. The \emph{overestimation} of anonymity set loss is favorable from the point of view of mixing users, who will, in reality, obtain higher privacy than implied by our conservative algorithm.

\subsubsection{Deployment setup}
Our analyses were performed on a machine with two 64-core AMD EPYC 7713 2.0 GHz processors, providing 256 threads in total, 2TB DDR4 3200 Mhz RAM (typically around 400GB was utilized), and Red Hat Enterprise Linux host OS, using Bitcoin blockchain state till date 2025-04-06 (block height 891137). As the direct compilation of the original code for a current OS is difficult due to dependency on outdated packages, we opted for a Podman-based containerization using a Ubuntu 20.04 image. 

\subsection{Results for  Whirlpool, Wasabi 1.x and Wasabi 2.x}
\label{sect:degradation_results}

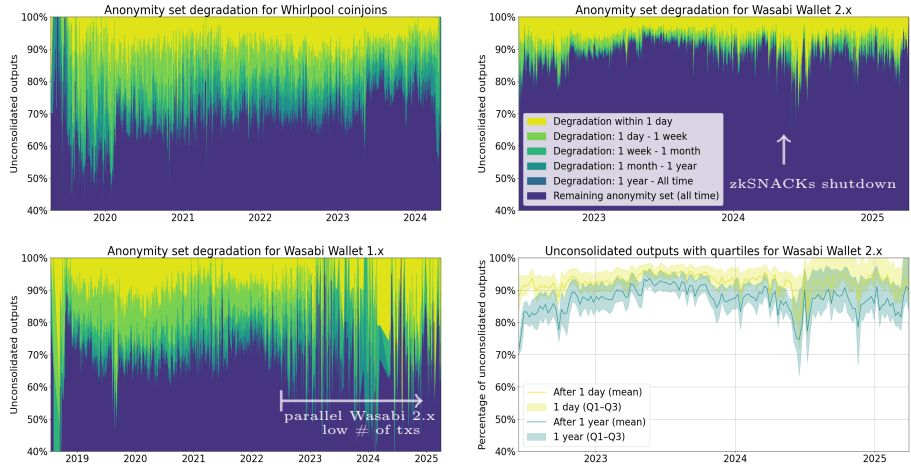
\begin{figure}[htbp]
    \centering
    \begin{minipage}[b]{0.49\textwidth}
    \centering
        \begin{overpic}[width=\linewidth,height=90px]{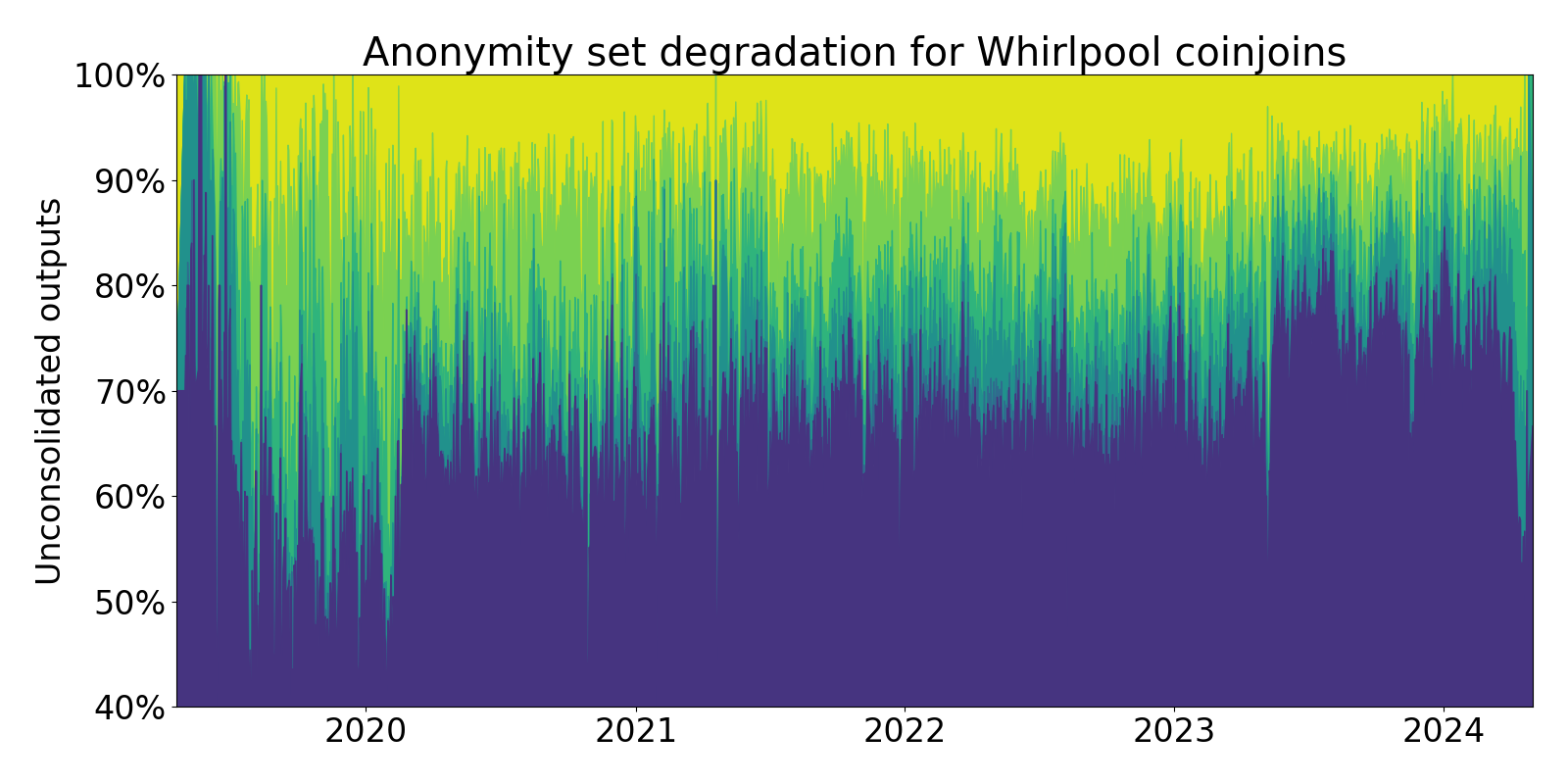}
        \end{overpic}
    \end{minipage}
\hfill
    \begin{minipage}[b]{0.49\textwidth}
        \centering
        \begin{overpic}[width=\linewidth,height=90px]{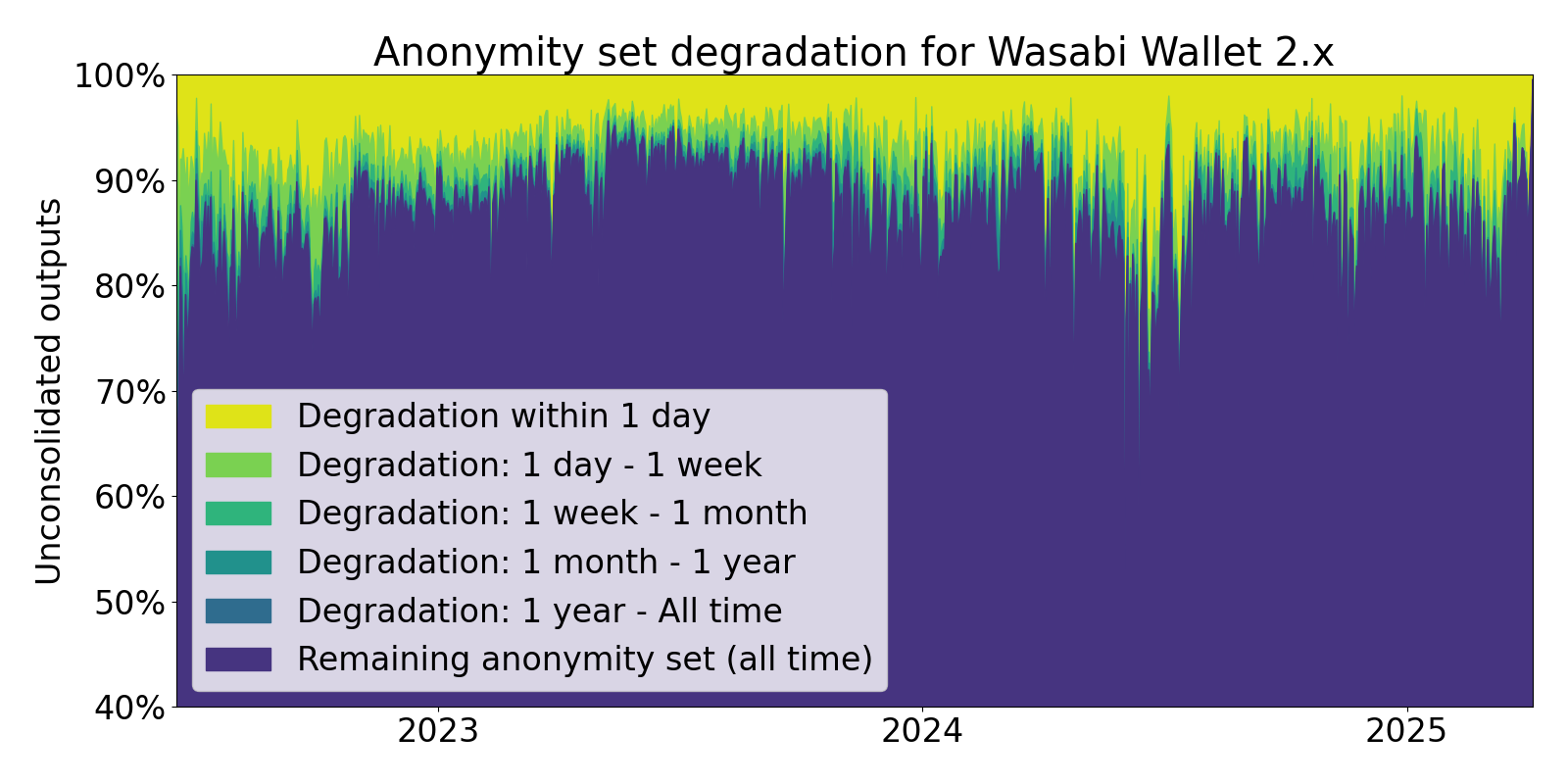}
            \put(58,10){\color{white}\tiny zkSNACKs shutdown}
            \put(68,15){
            \begin{tikzpicture}
                \draw[white, solid, line width=1pt, opacity=0.7, ->] (0,0) -- (0,0.4);
            \end{tikzpicture}}
        \end{overpic}
    \end{minipage}
\vspace{0.5cm}
    \begin{minipage}[b]{0.49\textwidth}
        \centering
        \begin{overpic}[width=\linewidth,height=90px]{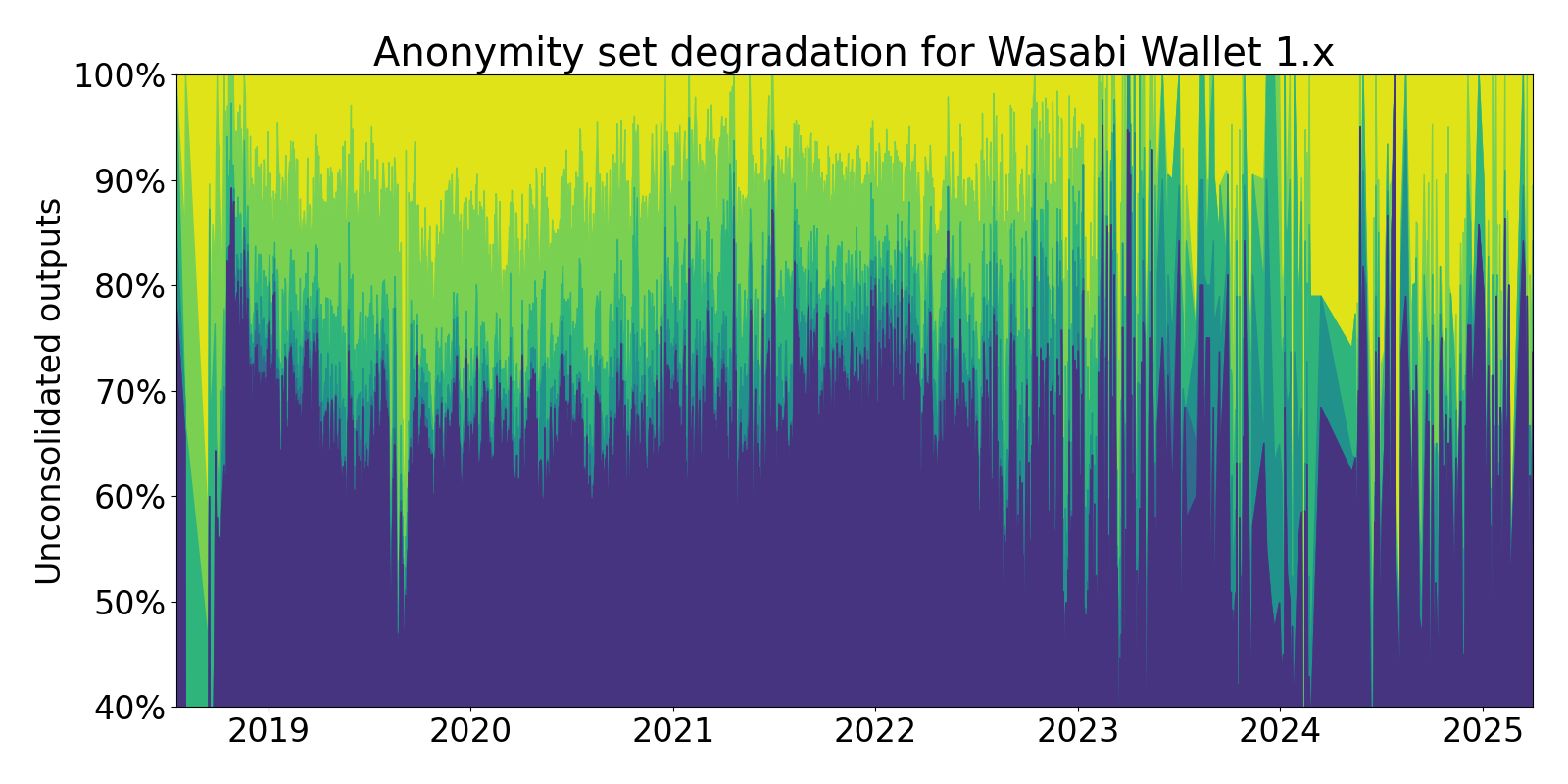}
            \put(63,12){\color{white}\tiny parallel Wasabi 2.x}
            \put(71.5,9){\color{white}\tiny low \# of txs}
            \put(60,16){
            \begin{tikzpicture}
                \draw[white, line width=1pt, opacity=0.7,  |->] (0,0) -- (1.9,0);          
            \end{tikzpicture}}
        \end{overpic}        
    \end{minipage}
\hfill
    \begin{minipage}[b]{0.49\textwidth}    
        \centering
        \begin{overpic}[width=\linewidth,height=90px]{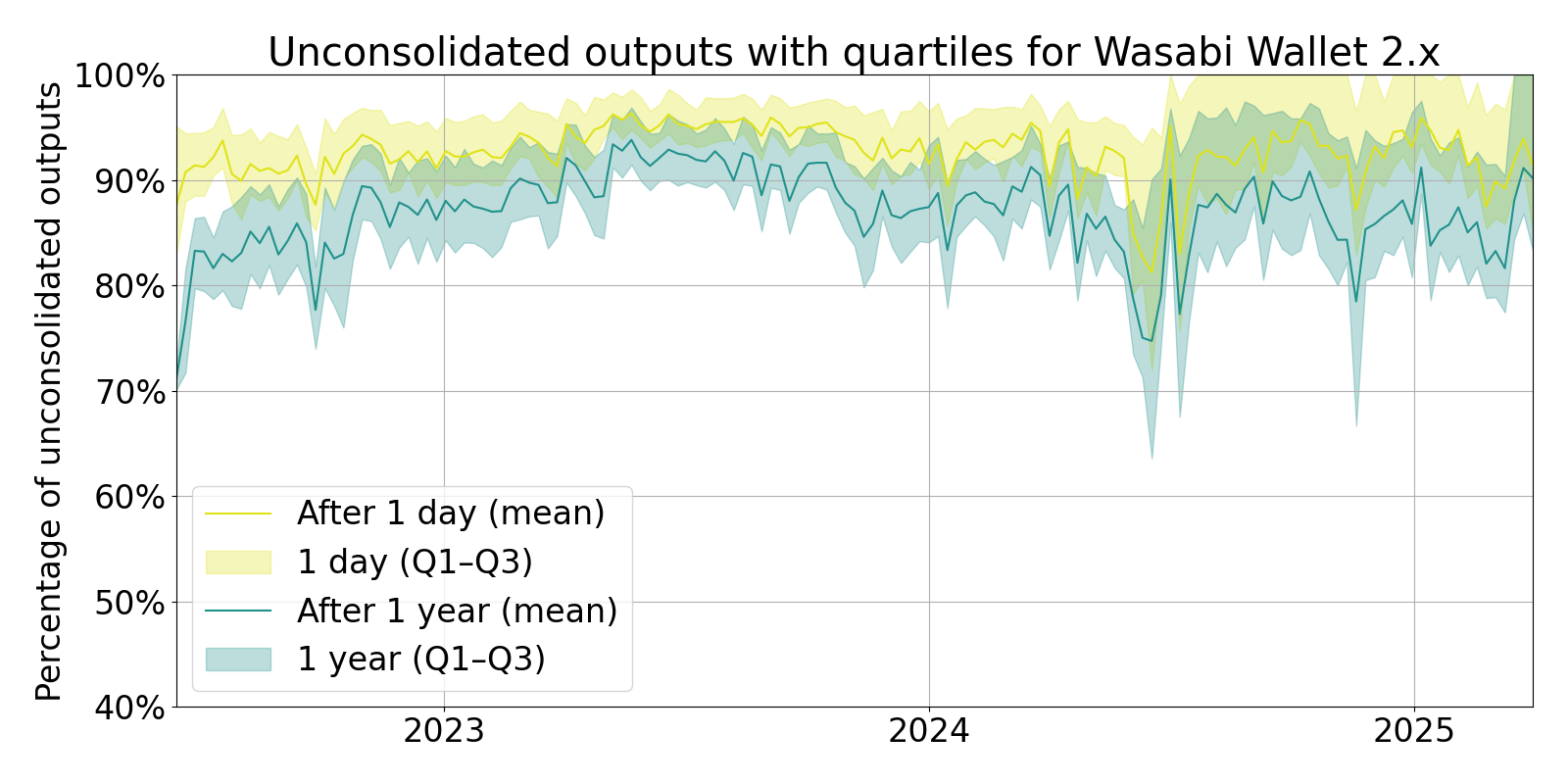}
        \end{overpic}
    \end{minipage}

    \caption{Anonymity set loss over time for Whirlpool, Wasabi 1.x, and Wasabi 2.x as a result of users' post-mix actions within a specific time interval. The darkest part marks the remaining fraction of the initial anonymity set. For Wasabi 2.x, the last sub-figure also shows the Q1 and Q3 quartiles. The anonymity set loss is averaged over all output denominations of coinjoins created at a given time. } 
    \label{fig:ww2degradation}
\end{figure}

We computed $A_d$ for days $d \in \{1, 7, 31, 365, \infty\}$ using \Cref{fig:entropy_loss_alg}. 
For all coinjoin designs, the biggest anonymity set degradation, caused by the post-mix behavior of users, usually happens within the very first day after the coinjoin creation, as visible in \Cref{fig:ww2degradation}. Conversely, any degradation after one year is usually negligible.  

For coinjoins with arbitrary values (Wasabi 1.1 and Wasabi 2.x), the anonymity set loss may differ across output denominations as discussed in \Cref{app:anon_loss_denoms}. The general trend is that smaller denominations are consolidated more frequently than the larger ones, with average results presented in \Cref{fig:ww2degradation} being close to smaller denominations due to their higher relative frequency in real coinjoins.

Averaging over all Whirlpool's pools, we observed a relatively stable $A_\infty$ -- between 30\% to 50\% (later coinjoins tend to have a lower loss),
as shown in the first sub-figure of \Cref{fig:ww2degradation}. The 0.001, 0.01, and 0.05 BTC pools exhibit a roughly comparable loss (\Cref{fig:swdegradation} in \Cref{app:anon_loss_denoms}). The largest 0.5 BTC pool has a smaller anonymity set loss overall, which might be explained by a lower need for any consolidation given an already large output.

For Wasabi 1.x, the anonymity set loss is around 25-40\% soon after its start until the first quarter of 2023, when it decreases significantly. However, the results from the later period are less representative as the number of Wasabi 1.x coinjoins significantly decreased from more than a thousand to smaller tens per month, in favor of the parallelly available operation of the superseding Wasabi 2.x design operated by the same coordinator. 

For Wasabi 2.x, the anonymity set loss is below 20\%, with a larger loss from June to August 2024 due to several contesting coordinators after the shutdown of the zkSNACKs coordinator. Afterwards, the \emph{kruw.io} coordinator became dominant, and its size of coinjoins, as well as the anonymity set loss returned to the values comparable to the (now defunct) zkSNACKs coordinator. The last sub-figure in \Cref{fig:ww2degradation} also shows first and third quartiles for anonymity set loss for Wasabi 2.x, exhibiting higher fluctuation for the post-zkSNACKs period, especially for the loss within a single day after a conjoin.  
Percentage-wise, the post-mix consolidations with a negative impact on the anonymity set loss happen significantly less for Wasabi 2.x due to the very high rate of remixes. Whirlpool and Wasabi 1.x have roughly comparable losses.

\section{Coinjoin mappings enumeration algorithm}
\label{sec:mappings_enum}

In this section, we discuss how to compute a coinjoin's privacy metrics in detail. The three main steps are as follows:
\begin{enumerate}
    \item enumerate all coinjoin mappings,
    \item compute the probability of each individual mapping,
    \item use results from steps 1 and 2 to compute privacy metrics for the whole coinjoin, as well as for individual users.
\end{enumerate}

\subsection{Enumeration of possible mappings}

The problem of enumerating all possible mappings of a coinjoin is super-exponential by its nature, as already enumerating all possible input partitions is super-exponential \cite{de2014asymptotic}. The algorithm that we use to enumerate the mappings consists of the following steps:
\begin{enumerate}
    \item fee preprocessing,
    \item additional knowledge preprocessing,
    \item computation of sub-mappings,
    \item sub-mapping filtering,
    \item enumeration of valid sub-mapping combinations into mappings.
\end{enumerate}

First, we describe the general parallelizable algorithmic approach for steps 3 and 5 without considering fees. 
Then, we discuss how to implement Step 1 for different fee structures and the changes needed in Step 3 to accommodate for fees. Next, we explore specific implementation properties to filter out sub-mappings (Step 4) that cannot occur. Finally, we describe how to use additional knowledge (Step 2) that an attacker can potentially have. In \Cref{app:multiple_cj}, we outline how to generalize this approach to multiple consecutive coinjoins.

\subsubsection{Algorithmic approach}
The core of our mapping enumeration is the computation of sub-mappings and their combination into mappings.
The computation of sub-mappings can be transformed into computing all solutions of a subset sum problem (SSP) \cite{ssp}. In SSP for a given set of values $S$ and a target value $t$, the goal is to find a subset of $S$ that sums to $t$. Finding sub-mappings of a coinjoin without fees can be transformed to an SSP instance by setting $t = 0$ and using the inputs and negated outputs as the elements of $S$. To solve the SSP, we opt for a backtracking approach. Similarly, we use backtracking to find compatible combinations of sub-mappings. Notice that backtracking is easily parallelizable, allowing for a fast multi-core implementation. Since we need to precompute all the sub-mappings, the worst-case memory complexity is exponential.

A naive approach is to enumerate all the mappings. However, coinjoins typically contain groups of the same-valued inputs and outputs. Therefore, such an enumeration would lead to a large number of mappings that are distinct only by a permutation of coins of the same value. To use this knowledge, we can enumerate only mappings up to a permutation of the same-valued coins. We call such mappings \emph{numeric}. It is possible to compute all the mappings corresponding to a numeric one. However, typically, it is sufficient to know just the number of the corresponding mappings for the computation of coinjoin-relevant privacy metrics. The number of numeric mappings, therefore, gives us a lower bound on the complexity of mapping-enumeration-based analysis of coinjoins.

\subsubsection{Fees}
For each fee structure described in \Cref{sec:background}, the enumeration algorithm needs a slightly different modification. 
If the coinjoin contains only predictable mining fees (Whirlpool) or predictable mining and coordination fees (Wasabi 1.x), we can remove the fees during transaction preprocessing and analyze the coinjoin as if it contained none. 

However, if there is an unpredictable component to any of the fees (e.g., decomposition fees in Wasabi 2.x), the enumeration needs to be modified. If we can estimate a lower bound on the fees, we can subtract it from the input and output values. Then, the difference $\delta$ between the upper and lower bound on the fees is a tolerance that needs to be added to the enumeration algorithm. Any sub-mapping where the difference between inputs and outputs is lower than $\delta$ needs to be considered for the possible mappings. 

Similarly, if the transaction can contain fees paid between users (JoinMarket), the sum of inputs and outputs of sub-mappings does not need to be exactly equal. In this case, the difference between inputs and outputs can be negative if the user gets paid to provide his funds for mixing. The number of sub-mappings with a positive or negative input-output difference can be considered when combining sub-mappings into mappings, as there could be a constraint on this property, e.g., JoinMarket has at most one sub-mapping with a positive difference. 

\subsubsection{Implementation restrictions on mappings}
If we assume that all the parties participating in the creation of a coinjoin use an unmodified implementation of the client wallet for the specific coinjoin protocol, we can limit the space of possible mappings based on the wallet's behavior. 

The coinjoins often follow some structure, e.g., Whirlpool allows only a fixed number of inputs and outputs in a transaction, and each user with a single client wallet has exactly one input and output. In other cases, the rules are more relaxed, but often, there is some limit on the number of inputs and outputs of each user or on the possible combinations of sub-mappings into mappings. 

These limitations can typically be adhered to by filtering the discovered sub-mappings. Alternatively, for faster computation, some of the rules can be implemented already as pruning rules in the backtracking search for sub-mappings. 

\subsubsection{Additional attacker's knowledge}
So far, the analysis uses solely information directly available from the coinjoin or the protocol used to create it, which can typically be assessed based on the structure of the transaction. However, the attacker can obtain additional information about the transaction's inputs and outputs as described in \Cref{sec:blocksci_coinjoinloss}.

If the attacker knows that some inputs (or outputs) belong to the same user, during the preprocessing of the transaction, he can sum their values and merge them into a single input (output). When enumerating all mappings (not just numerical ones), the attacker can also use the information that some inputs or outputs belong to different users, filtering out any sub-mappings that attribute them to the same user. 

If the attacker knows that an input-output pair is owned by a single user, he can replace the higher one with the difference of their values and remove the lower one. 
These operations reduce the coinjoin size, making it easier to analyze.

\subsection{Mapping probability}
To estimate each user's privacy in a coinjoin, we need to know how likely individual mappings are to occur. This depends on multiple factors: 
\begin{itemize}
    \item the distribution and privacy level of coins that the users want to mix, 
    \item the input selection of a coinjoin,
    \item the output selection of a coinjoin.
\end{itemize}
We will take a Bayesian standpoint, treating unknowns as random variables. Then we can compute the probability $p(M)$ of the mapping $M$ occurring as $p(M) = p(M | A) \cdot p(A)$, where $A: I \rightarrow U$ assigns the inputs to their owners.

The probability $p(A)$ depends on the first two out of the three factors above. In some cases, this probability is fully determined by the input coin selection implemented in the wallet software. For example, the Samourai wallet (for Whirlpool coinjoin) always selects just a single input regardless of the number of coins in the wallet. The opposite extreme would be a wallet always registering all its coins to a coinjoin; $p(A)$ would then be fully determined by the probability distribution of coins among users. 

The probability $p(M | A)$ depends solely on the wallet's output selection implementation. In many cases, this process is deterministic; therefore, given the input attribution to users, all but one numeric mapping has a probability of zero. However, some implementations, including Wasabi 2.x, randomize this process, making the analysis more difficult as discussed in \Cref{sec:ww2_probability}.

\subsection{New privacy metric}

Previous works \cite{boltzmann,maurer2017anonymous} computed the entropy as $E(T) = \log_2(|\mathcal{M}_T|)$, a special case of the Shannon entropy when all the mappings have the same probability $1/|\mathcal{M}_T|$ of occurring. We consider the more general case
$$E(T) \coloneqq -\sum_{M\in \mathcal{M}_T}p(M)\cdot \log(p(M)).$$

This entropy captures how unpredictable the whole coinjoin is, and allows comparing different coinjoin designs. However, from the user's perspective, it is more important to asses the unpredictability of their specific sub-mapping. We could directly compute the probability of the user's actual sub-mapping $S$ as $$p(S) \coloneqq \sum_{M \in \{M' \in \mathcal{M} | S \in M'\}} p(M).$$

In practice, though, the anonymity of different outputs within a sub-mapping can vary dramatically (e.g., depending on the number of times the same value is present), so $p(S)$ might not be a fine enough metric.

The usual measure adopted by previous work \cite{boltzmann,maurer2017anonymous} is the probability $p(i,o)$ that a given input-output pair is linked, with the additional information about mappings' probability computed as
$$p(i,o) \coloneqq \sum_{M \in \{M' \in \mathcal{M} | \exists (I, O) \in M': i \in I, o \in O\}} p(M).$$ 
This can be useful for analysing the transaction, but not so much for a regular user.
We propose a new metric providing a conservative estimate of how well a user's output $o$ is mixed, using the knowledge of their set of inputs $I$. We compute it as $$p(I, o) \coloneqq \max_{i \in I} p(i, o).$$
Clearly $p(i', o) \leq p(I, o)$ for any $i'\in I$, so if an attacker finds out that this user owns $i'$, our metric bounds the information gained. While computing $p(M)$ is typically still not feasible for real-world coinjoins, it entails a more detailed description than the practically used metrics based on anonymity sets.

\section{Evaluation of Wasabi 2.x coinjoins}
\label{sect:ww2_mapping_enums}
In this section, we evaluate our approach on coinjoins created by Wasabi 2.x. We chose Wasabi 2.x over Wasabi 1.x and Samourai Whirlpool as these two are no longer active (shutdown in 2024) and have very strict limitations on sub-mappings, making the enumeration trivial. While JoinMarket is actively used and allows non-trivial mappings, the taker in this protocol has complete control over the final transaction, allowing him to estimate his privacy better. Thus, we preferred Wasabi 2.x, where no single entity knows the complete mapping.

We also employ the anonymity set loss observed for real coinjoins in \Cref{sec:blocksci_coinjoinloss} as one of the inputs for practical computations of the mapping enumeration. Firstly, we utilize concrete consolidations for a given coinjoin to decrease the effective number of outputs by merging ones consolidated under the same transaction into a single one. Secondly, we use the average anonymity set loss to lower the expected size of larger coinjoins during complexity extrapolation.

\subsection{Mapping enumeration}
\label{sect:mappings_enum}
The necessary modifications of the enumeration algorithm for Wasabi 2.x are described in \Cref{app:WW2_mappings}. 
To obtain data for the evaluation, we used a Wasabi 2.x emulator \cite{emucj} with coins randomly assigned to wallets so that the resulting transactions have a small number of inputs and outputs. Emulations provide us with a large number of small coinjoins where we know the ground truth mappings, allowing us to validate our approach by checking that the emulated mapping is among the enumerated numeric mappings. 

\begin{figure}
\centering
\begin{minipage}[t]{.6\textwidth}
    \includegraphics[width=1\linewidth, left]{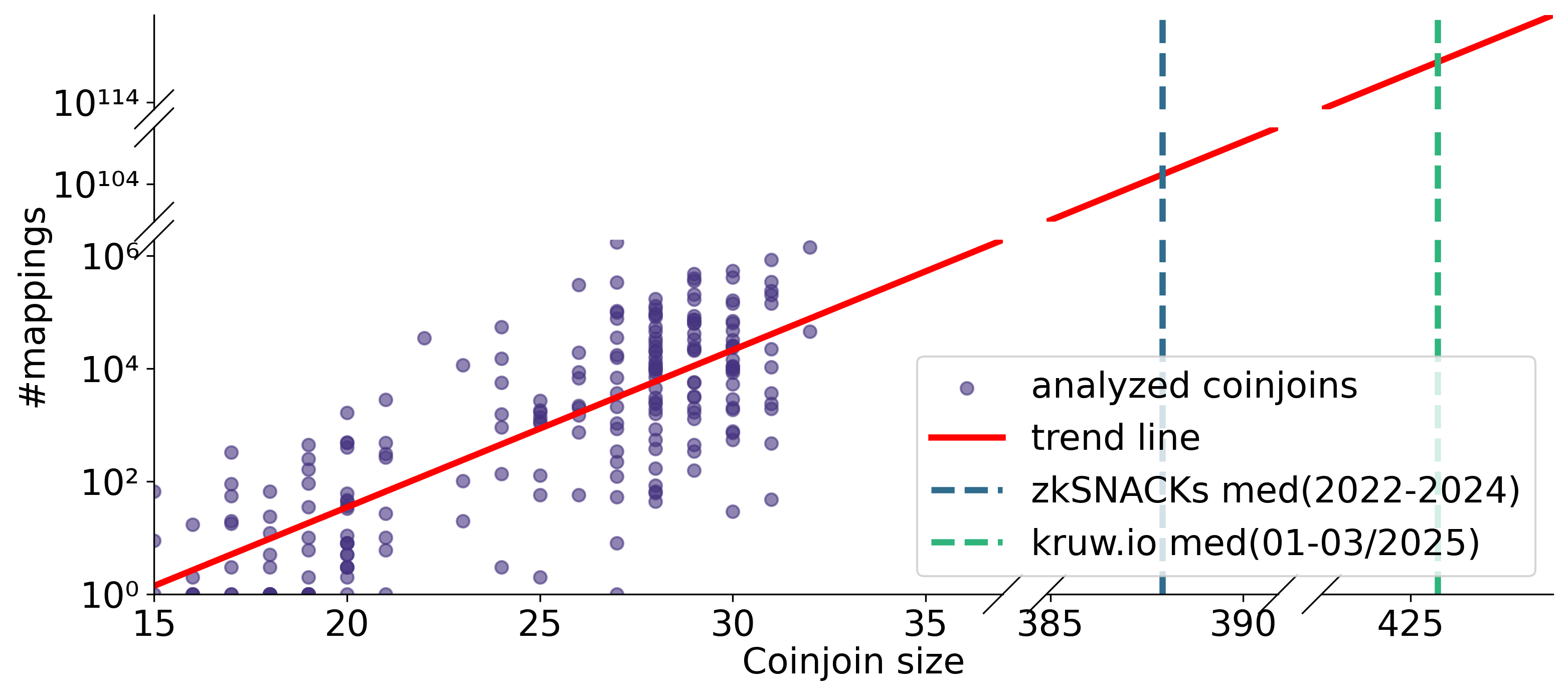}
    \captionof{figure}{Number of numeric mappings for different sizes of coinjoins. The size is defined as the sum of the number of inputs and outputs. The trend line shows how the number of numeric mappings grows with size. The y-axis is logarithmic.}
    \label{fig:enumeration}
\end{minipage}%
\hfill
\begin{minipage}[t]{.35\textwidth}
    \includegraphics[width=1\linewidth, right]{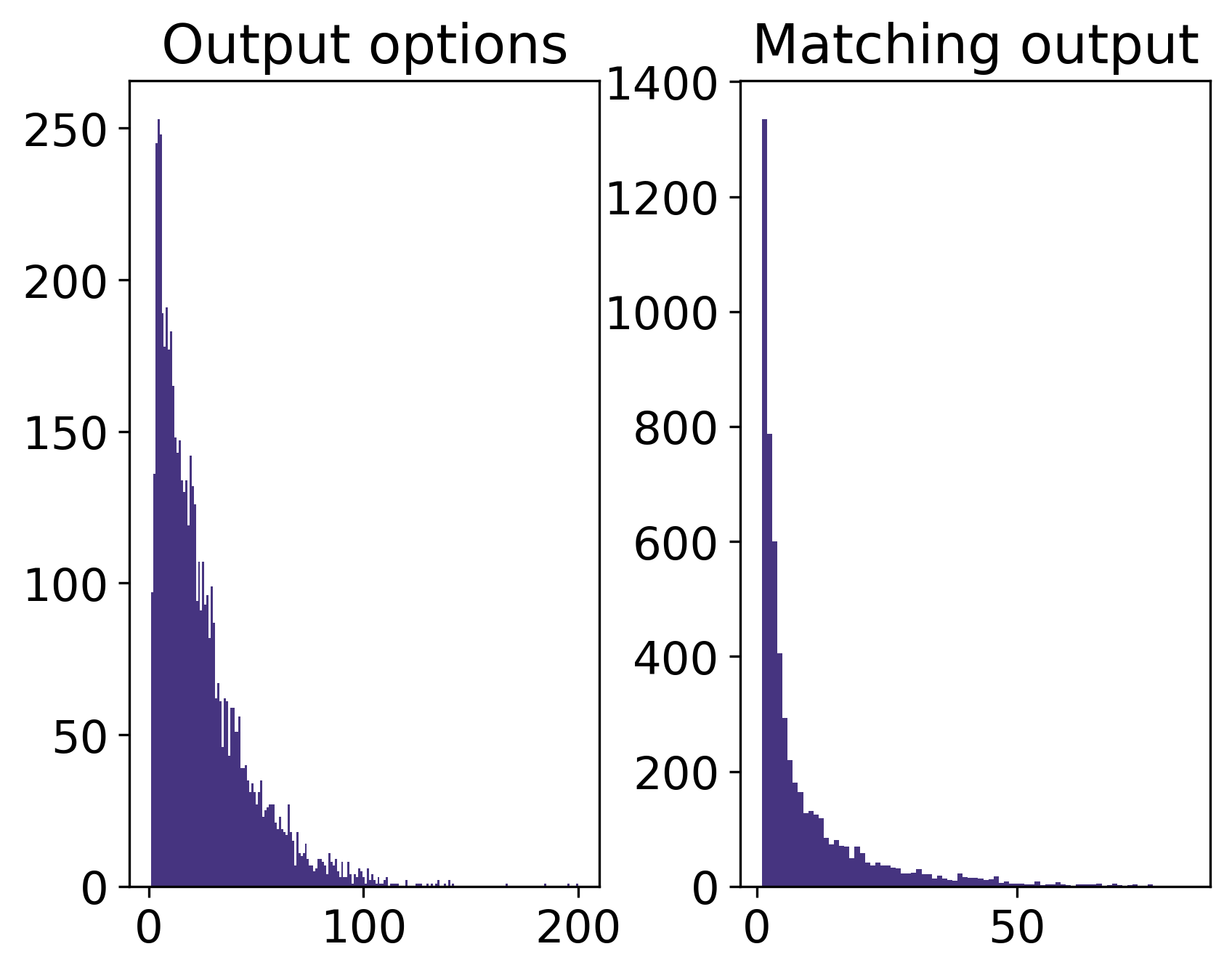}
    \captionof{figure}{Distribution of the number of possible output selections and the order of the matching output.}
    \label{fig:sake}
\end{minipage}
\end{figure}

We evaluated 220 transactions in total. In all cases, the real mapping from the emulation was among the enumerated mappings, indicating that our method does not omit possible mappings. \Cref{fig:enumeration} shows the number of mappings for different coinjoin sizes, where the size is defined as the sum of the number of inputs and outputs. 

While the variance in the number of numeric mappings for the same size is high, we can see a clear exponential growth for the average case. We can extrapolate this trend to larger coinjoin sizes to evaluate the expected number of numeric mappings for real-world coinjoins. The graph shows the expected average number of numeric mappings for a median coinjoin produced under the zkSNACKs coordinator (2022-2024) and the kruw.io coordinator in 2025. Their size is lowered by the percentage of consolidations observed in \Cref{sec:blocksci_coinjoinloss}. In both cases, the enumeration of all numeric mappings would be infeasible, suggesting that users get sufficient privacy unless an attacker obtains additional information to be able to analyze the transaction properly.

We further found five small real Wasabi 2.x coinjoins and enumerated the possible numeric mappings for them. For real transactions, we were able to use the output consolidations. However, unlike in the \Cref{sec:blocksci_coinjoinloss}, we linked together only the outputs that are almost certainly owned by a single entity to avoid potentially falsely assumed consolidations. A table with the transaction IDs and their number of numeric mappings is available in \Cref{app:real_cjs}.

\subsection{Mapping probability}
\label{sec:ww2_probability}
Computing $p(M)$ consists of computing $p(A)$ and $p(M | A)$ separately.
We do not know $A$, but from a privacy perspective, it is better to assume that the attacker knows $p(A)$, or even worse, knows which assignment $A$ really occurred. 
With such an assumption, $p(M)$ depends only on the output selection of Wasabi 2.x. A direct way to obtain the probability of a coinjoin mapping would be to repeatedly simulate the output selection for all the wallets in the given coinjoin. Then, from the observed results, we would filter the mappings that have the same resulting set of output values as the original coinjoin, and assess their probability based on the number of their occurrences. However, such a simulation would likely produce an extreme number of different outcomes, making it infeasible to simulate the process enough times to obtain the probabilities. Therefore, we propose to simulate wallets individually and assign probabilities to sub-mappings, which is more efficient. Then, based on the probabilities of sub-mappings, compute the probability of a whole mapping. 

To provide statistics on the output selection, we used 100 emulated coinjoins of realistic size -- with hundreds of inputs and outputs. Then, we modified the Wasabi 2.x output-selection simulator Sake \cite{sake} and used it to simulate the output selection. When simulating all the wallets, we obtained the exact matching sub-mapping on the first try for 29\% of the wallets, showing relatively low entropy in the output selection. Furthermore, we modified the simulator to return all the sub-mappings for a given user, from which it selects randomly in the last step of output selection. We ran it 100 times for each wallet, as the set of possible outputs can differ. \Cref{fig:sake} shows the distribution of the number of obtained sub-mappings for individual wallets and the distribution of indices of matching decompositions when ordered by their estimated probability. 
In 3\% of cases, the emulated mapping did not occur among the simulated ones, suggesting that either more iterations would be needed or that the Sake simulator and the real Wasabi 2.x wallets behave differently.
An additional mismatch between Sake and Wasabi 2.x led us to discover bugs in the Wasabi 2.x described in \Cref{app:ww2_bugs}.

\section{Conclusions}
\label{sect:conclusions}

The paper tackles an open question of accurate anonymity estimation of privacy-enhancing trustless coinjoins with a centralized coordinator. We first show that despite frequent post-mix consolidations by participating users, the anonymity set loss for coinjoin outputs stabilizes and does not deteriorate significantly after the first year of coinjoin creation. Together with an exact fee computation and implementation-specific limits, we propose an improved algorithm to enumerate input-output mappings suitable especially for Wasabi 2.x coinjoins, practical enough to fully enumerate real coinjoins of smaller coordinators and provide extrapolated complexity estimations for the large ones. We conclude that large real Wasabi 2.x coinjoins still provide sufficient privacy even against improved analytical methods utilizing information extractable from the public blockchain.

\vspace{0.2cm}
\noindent \textbf{Acknowledgments:} The work was supported by EU CHESS grant \#101087529.

This preprint has not undergone any post-submission improvements or corrections. The Version of Record of this contribution is published in Computer Security – ESORICS 2025 \href{https://doi.org/10.1007/978-3-032-07901-5_7}{https://doi.org/10.1007/978-3-032-07901-5\_7}.

%
%

 \bibliographystyle{splncs04}
 \bibliography{bibliography}

\begin{thebibliography}{10}
\providecommand{\url}[1]{\texttt{#1}}
\providecommand{\urlprefix}{URL }
\providecommand{\doi}[1]{https://doi.org/#1}

\bibitem{lnk_androulaki}
Androulaki, E., Karame, G.O., Roeschlin, M., Scherer, T., Capkun, S.:
  {{Evaluating User Privacy in Bitcoin}}. In: Financial Cryptography and Data
  Security. pp. 34--51. Springer Berlin Heidelberg (2013)

\bibitem{boltzmann}
{Boltzmann authors}: Boltzmann (2016),
  \url{https://github.com/Samourai-Wallet/boltzmann}, visited on 2025-04-28

\bibitem{btcpay}
{BTCPay Server}: {BTCPay Server}, \url{https://btcpayserver.org/}, visited on
  2025-04-28

\bibitem{de2014asymptotic}
De~Bruijn, N.G.: Asymptotic methods in analysis. Courier Corporation (2014)

\bibitem{mapreduce}
Dean, J., Ghemawat, S.: {MapReduce: simplified data processing on large
  clusters}. Commun. ACM  \textbf{51}(1),  107–113 (Jan 2008)

\bibitem{emucj}
Dufka, A., Rýpar, D.: {EmuCoinJoin: A container-based setup for the emulation
  of CoinJoin transactions on RegTest network},
  \url{https://github.com/crocs-muni/coinjoin-emulator}, visited on 2025-11-03

\bibitem{dumplings}
Ficsor, A.: Dumplings (2020), \url{https://github.com/nopara73/Dumplings},
  visited on 2025-04-28

\bibitem{sake}
Ficsor, A.: Sake (2021), \url{https://github.com/nopara73/Sake}, visited on
  2025-04-28

\bibitem{zerolink}
Ficsor, A.: {ZeroLink: The Bitcoin Fungibility Framework} (2022),
  \url{https://nopara73.medium.com/introducing-zerolink-the-bitcoin-fungibility-framework-dc5338086198},
  visited on 2025-04-28

\bibitem{ficsor2021wabisabi}
Ficsor, A., Seres, I.A., Kogman, Y., Ontivero, L.: {Wabisabi: Centrally
  coordinated coinjoins with variable amounts} (2021)

\bibitem{DBLP:journals/popets/GoldfederKRN18}
Goldfeder, S., Kalodner, H.A., Reisman, D., Narayanan, A.: {When the cookie
  meets the blockchain: Privacy risks of web payments via cryptocurrencies}.
  Proc. Priv. Enhancing Technol.  \textbf{2018}(4),  179--199 (2018)

\bibitem{jm}
{JoinMarket authors}: {JoinMarket},
  \url{https://github.com/JoinMarket-Org/joinmarket-clientserver}, visited on
  2025-04-28

\bibitem{blocksci20}
Kalodner, H., M{\"o}ser, M., Lee, K., Goldfeder, S., Plattner, M., Chator, A.,
  Narayanan, A.: {{B}lock{S}ci: Design and applications of a blockchain
  analysis platform}. In: USENIX Security '20. pp. 2721--2738 (Aug 2020)

\bibitem{ssp}
Kellerer, H., Pferschy, U., Pisinger, D.: {{The Subset Sum Problem}}, pp.
  73--115. Springer Berlin Heidelberg (2004)

\bibitem{lnk_maesa2016uncovering}
Maesa, D.D.F., Marino, A., Ricci, L.: Uncovering the bitcoin blockchain: an
  analysis of the full users graph. In: {IEEE International Conference on Data
  Science and Advanced Analytics (DSAA)}. pp. 537--546 (2016)

\bibitem{maurer2017anonymous}
Maurer, F.K., Neudecker, T., Florian, M.: {Anonymous CoinJoin transactions with
  arbitrary values}. In: IEEE trustcom/bigdatase/icess. pp. 522--529 (2017)

\bibitem{maxwell_cj}
Maxwell, G.: {CoinJoin: Bitcoin privacy for the real world} (2013),
  \url{https://bitcointalk.org/index.php?topic=279249.0}, visited on 2025-04-28

\bibitem{meiklejohn2013fistful}
Meiklejohn, S., Pomarole, M., Jordan, G., Levchenko, K., McCoy, D., Voelker,
  G.M., Savage, S.: A fistful of bitcoins: characterizing payments among men
  with no names. In: Internet measurement conference. pp. 127--140 (2013)

\bibitem{joinmarket2017}
M{\"{o}}ser, M., B{\"{o}}hme, R.: The price of anonymity: empirical evidence
  from a market for bitcoin anonymization. J. Cybersecur.  \textbf{3}(2),
  127--135 (2017)

\bibitem{lnk_Reid2013}
Reid, F., Harrigan, M.: {An Analysis of Anonymity in the Bitcoin System}, pp.
  197--223. Springer New York (2013)

\bibitem{lnk_dorit}
Ron, D., Shamir, A.: {Quantitative Analysis of the Full Bitcoin Transaction
  Graph}. In: Financial Cryptography and Data Security. pp. 6--24. Springer
  Berlin Heidelberg (2013)

\bibitem{sw}
{Samourai authors}: {Samourai Wallet - Whirlpool},
  \url{https://web.archive.org/web/20240417214653/https://www.samouraiwallet.com/whirlpool},
  visited on 2025-04-28

\bibitem{schnoering2023heuristics}
Schnoering, H., Vazirgiannis, M.: Heuristics for detecting coinjoin
  transactions on the bitcoin blockchain. arXiv:2311.12491  (2023)

\bibitem{coinjoin_adoption_22}
St\"{u}tz, R., Stockinger, J., Moreno-Sanchez, P., Haslhofer, B., Maffei, M.:
  {Adoption and Actual Privacy of Decentralized CoinJoin Implementations in
  Bitcoin}. In: AFT '22. p. 254–267. ACM, New York, USA (2023)

\bibitem{detectWasabi22}
Tironsakkul, T., Maarek, M., Eross, A., Just, M.: {The Unique Dressing of
  Transactions: Wasabi CoinJoin Transaction Detection}. In: EICC '22. p.
  21–28. ACM, New York, USA (2022)

\bibitem{wahrstatter2024reducing}
Wahrst{\"a}tter, A., Taudes, A., Svetinovic, D.: {Reducing Privacy of CoinJoin
  Transactions: Quantitative Bitcoin Network Analysis}. IEEE Transactions on
  Dependable and Secure Computing  \textbf{21}(5) (2024)

\bibitem{ww2}
{Wasabi Wallet authors}: {Wasabi Wallet - Bitcoin privacy wallet with
  coinjoin}, \url{https://wasabiwallet.io/}, visited on 2025-04-28

\bibitem{ww1}
zkSNACKs: {Wasabi Wallet},
  \url{https://github.com/WalletWasabi/WalletWasabi/tree/v1.98.4.0}, visited on
  2025-04-28

\end{thebibliography}

 \appendix

 \section{Anonymity set loss results}
\label{app:anon_loss_denoms}

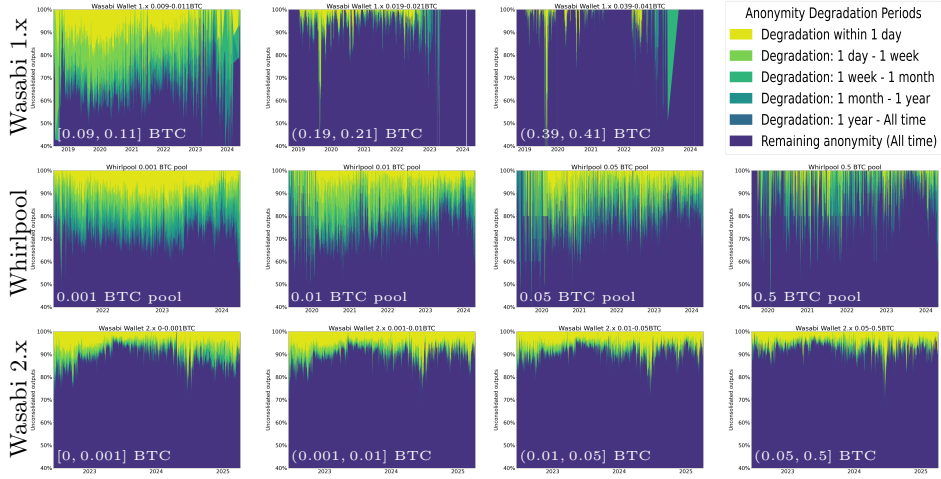
\begin{figure}[htbp]
    \centering

    \begin{minipage}[b]{0.24\textwidth}
    \centering
        \begin{overpic}[width=\linewidth,height=60px]{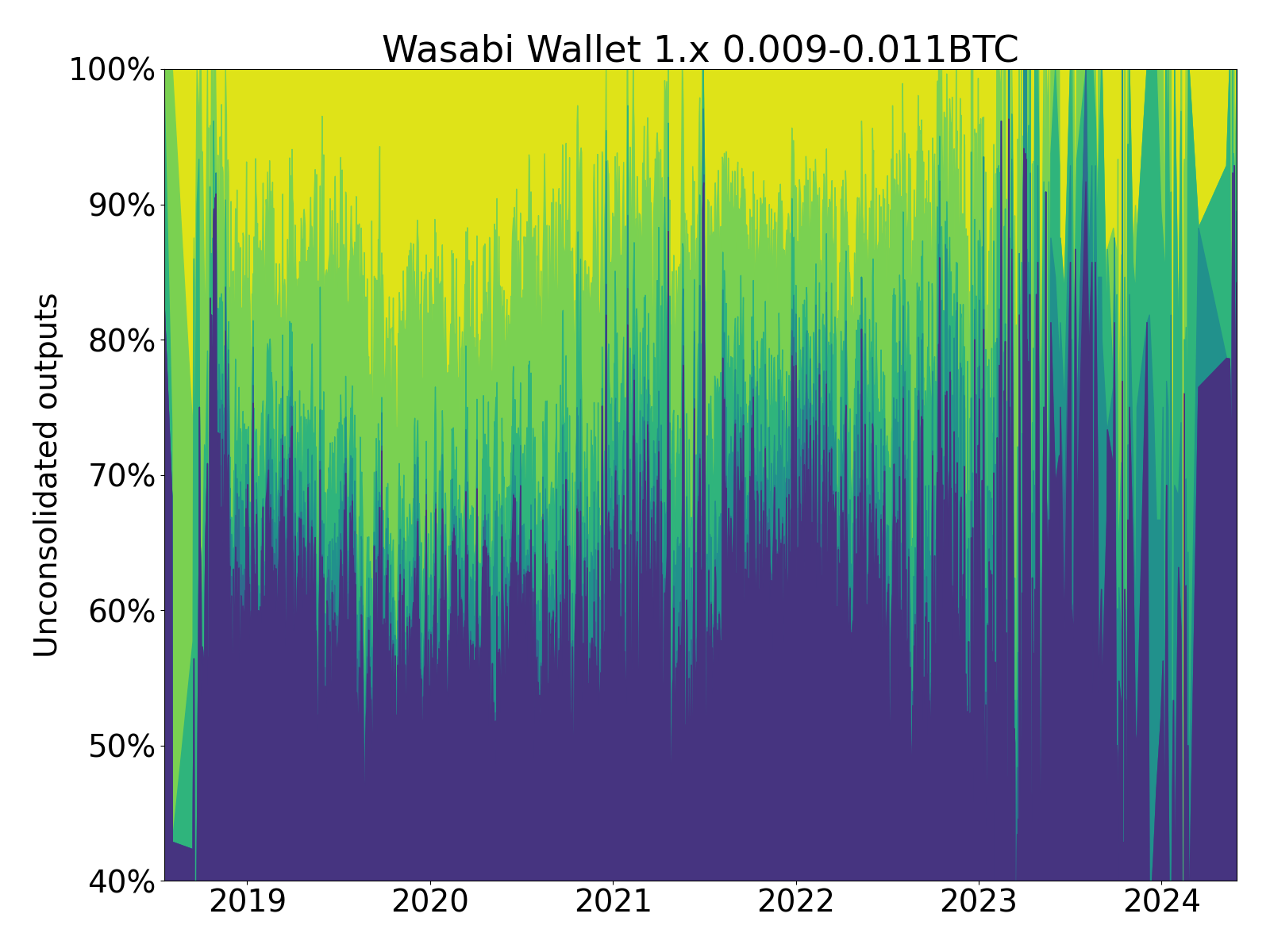}
        \put(14,9){\color{white}\tiny \( [0.09, 0.11] \) BTC}
         \put(-7,10){\rotatebox{90}{Wasabi 1.x}}
        \end{overpic}
    \end{minipage}
\hfill
    \begin{minipage}[b]{0.24\textwidth}
        \centering
        \begin{overpic}[width=\linewidth,height=60px]{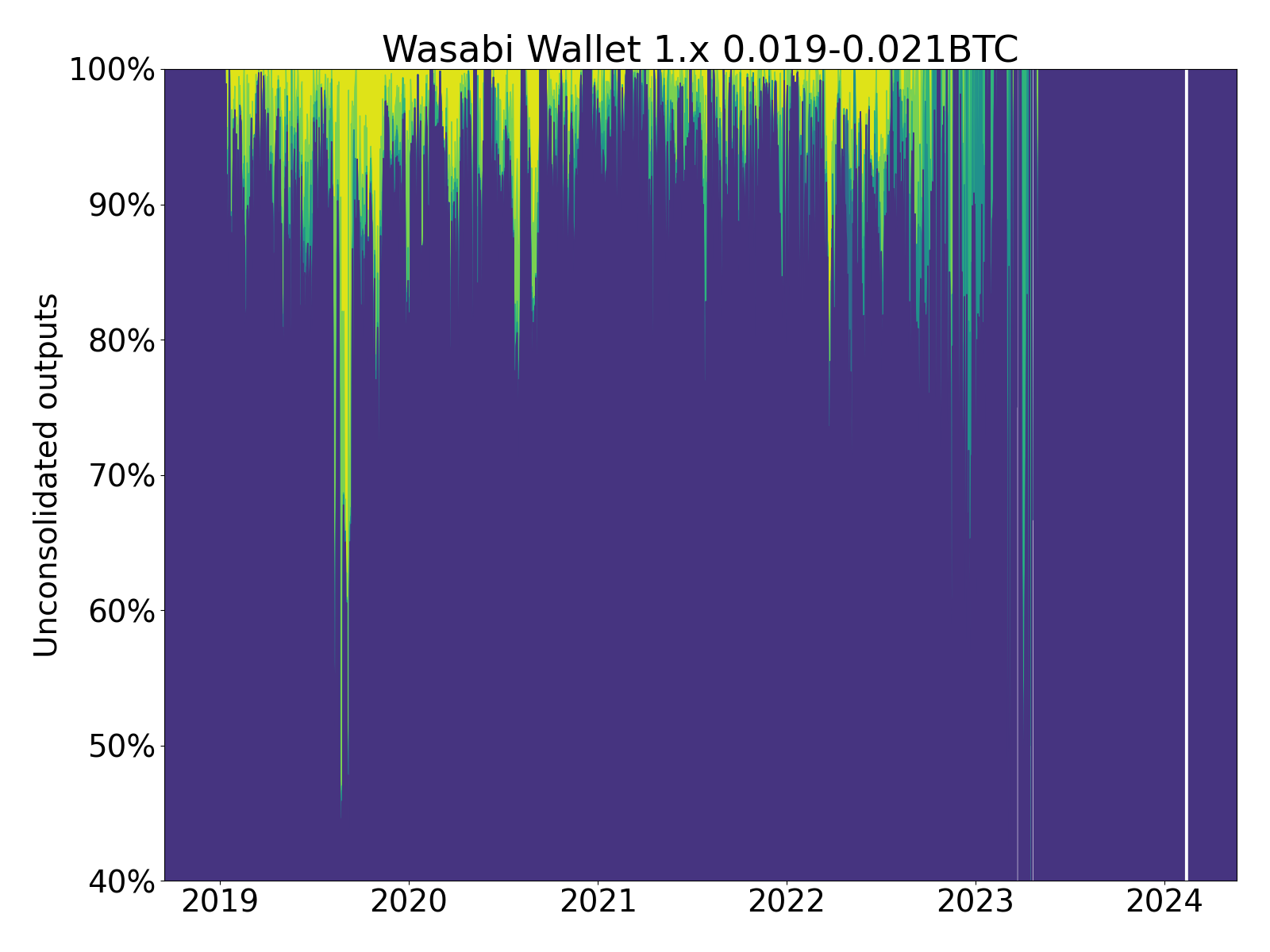}
        \put(14,9){\color{white}\tiny \( (0.19, 0.21] \) BTC}
        \end{overpic}
    \end{minipage}
    \begin{minipage}[b]{0.24\textwidth}
        \centering
        \begin{overpic}[width=\linewidth,height=60px]{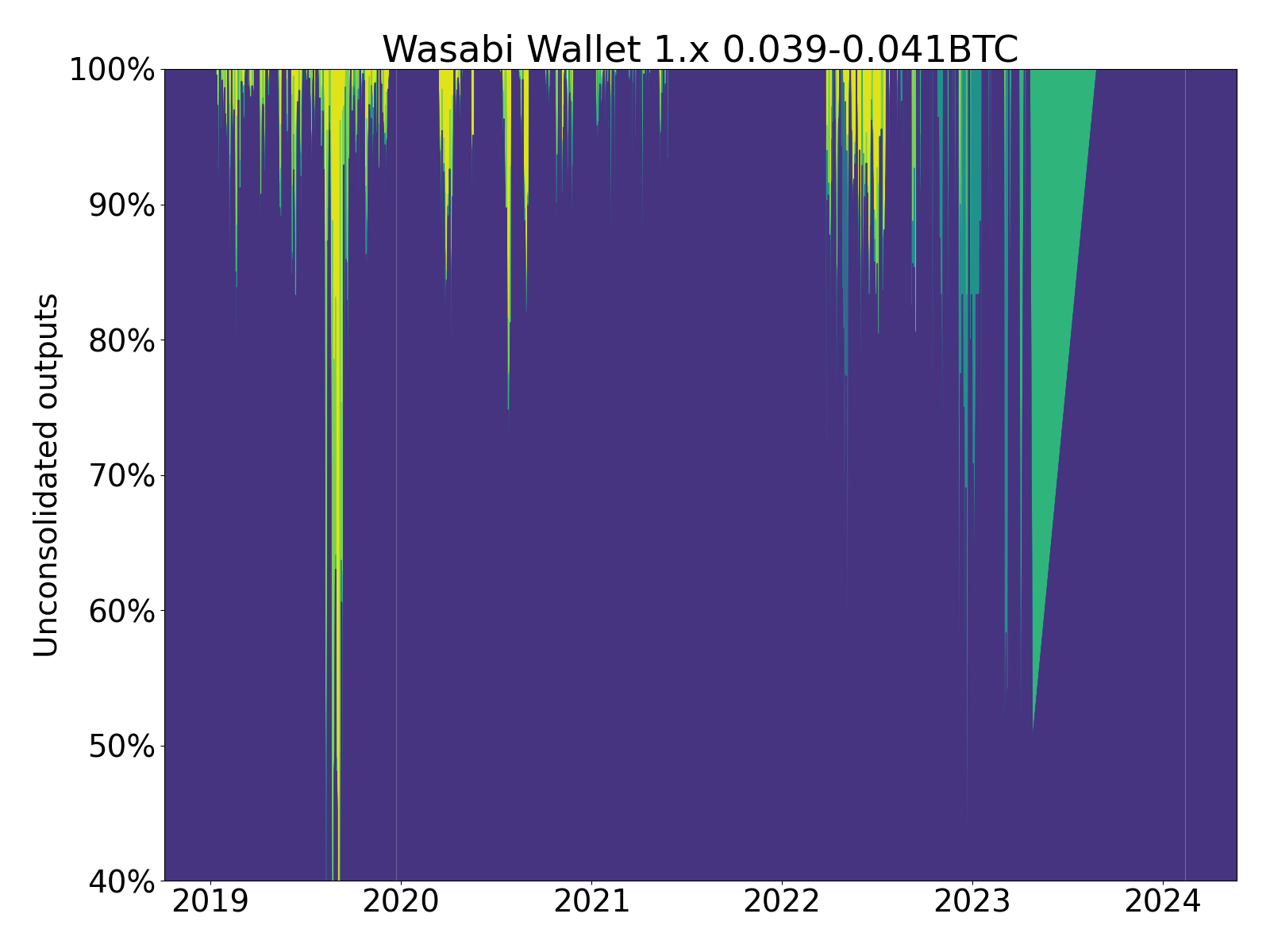}
        \put(14,9){\color{white}\tiny \( (0.39, 0.41] \) BTC}
        \end{overpic}        
    \end{minipage}
\hfill
    \begin{minipage}[b]{0.24\textwidth}    
        \centering
        \begin{overpic}[width=\linewidth,height=60px]{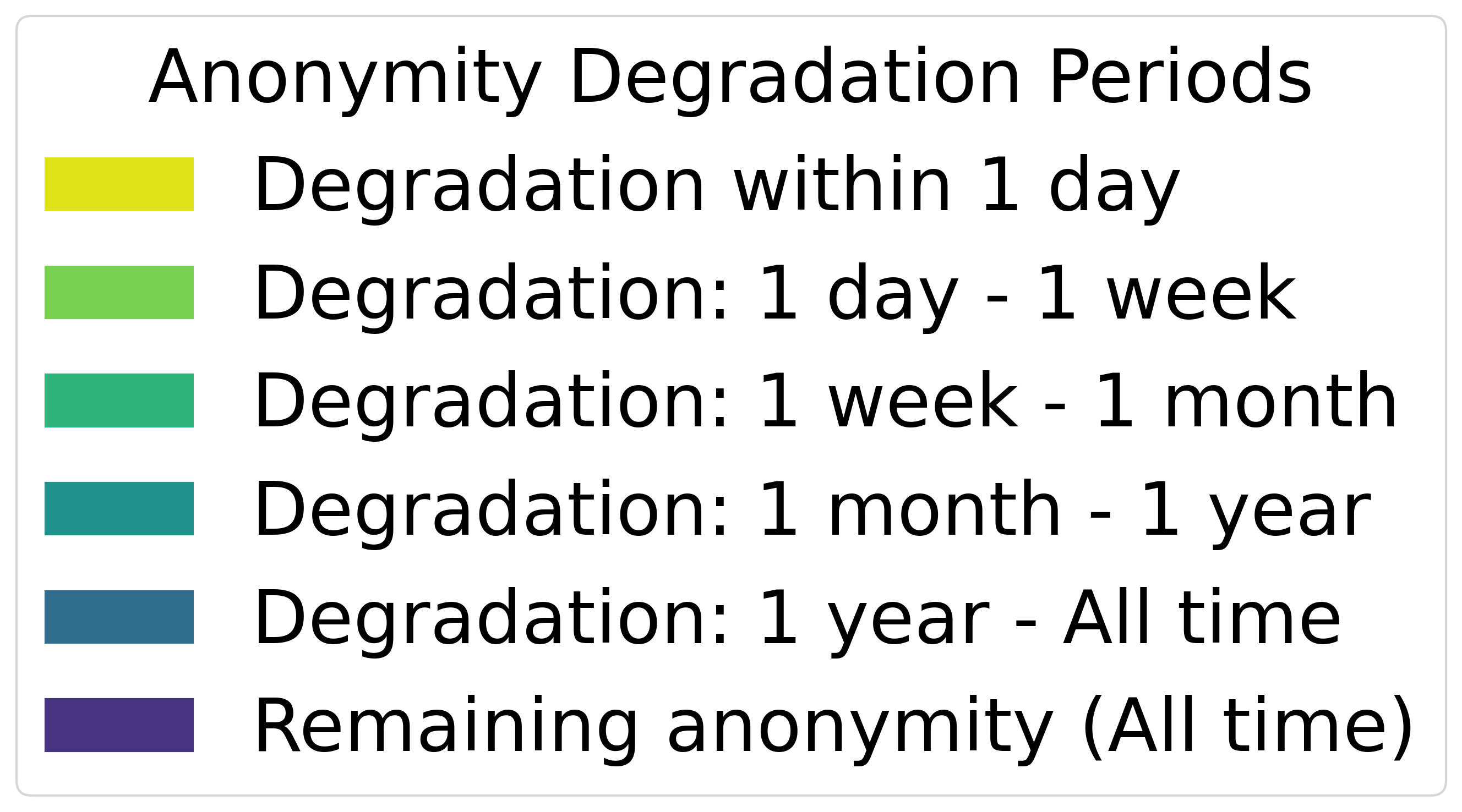}
        \put(14,9){\color{white}\tiny 0.5 BTC pool}
        \end{overpic}
    \end{minipage}   
    
    \begin{minipage}[b]{0.24\textwidth}
    \centering
        \begin{overpic}[width=\linewidth,height=60px]{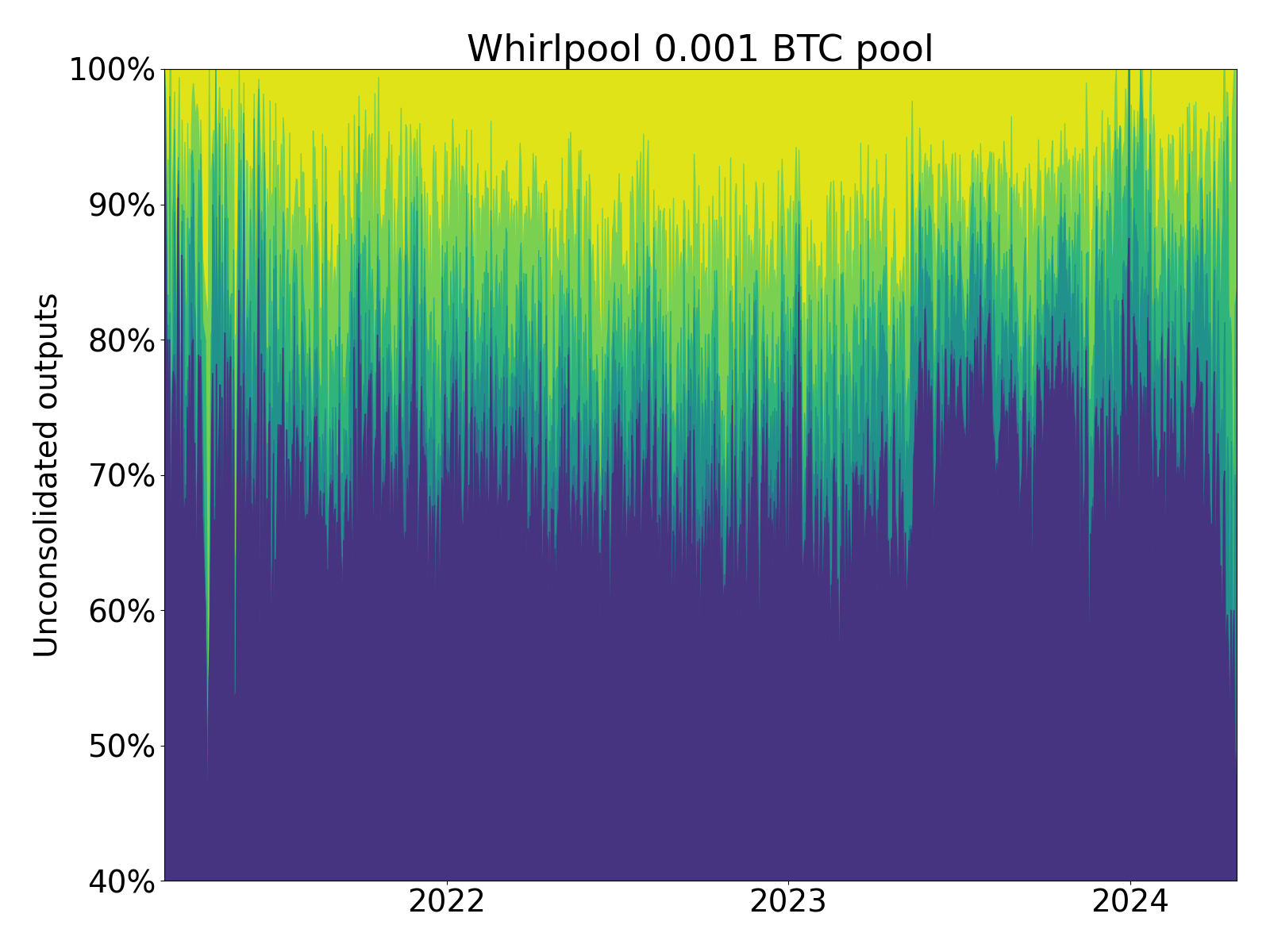}
        \put(14,8){\color{white}\tiny 0.001 BTC pool}
         \put(-7,10){\rotatebox{90}{Whirlpool}}
        \end{overpic}
    \end{minipage}
\hfill
    \begin{minipage}[b]{0.24\textwidth}
        \centering
        \begin{overpic}[width=\linewidth,height=60px]{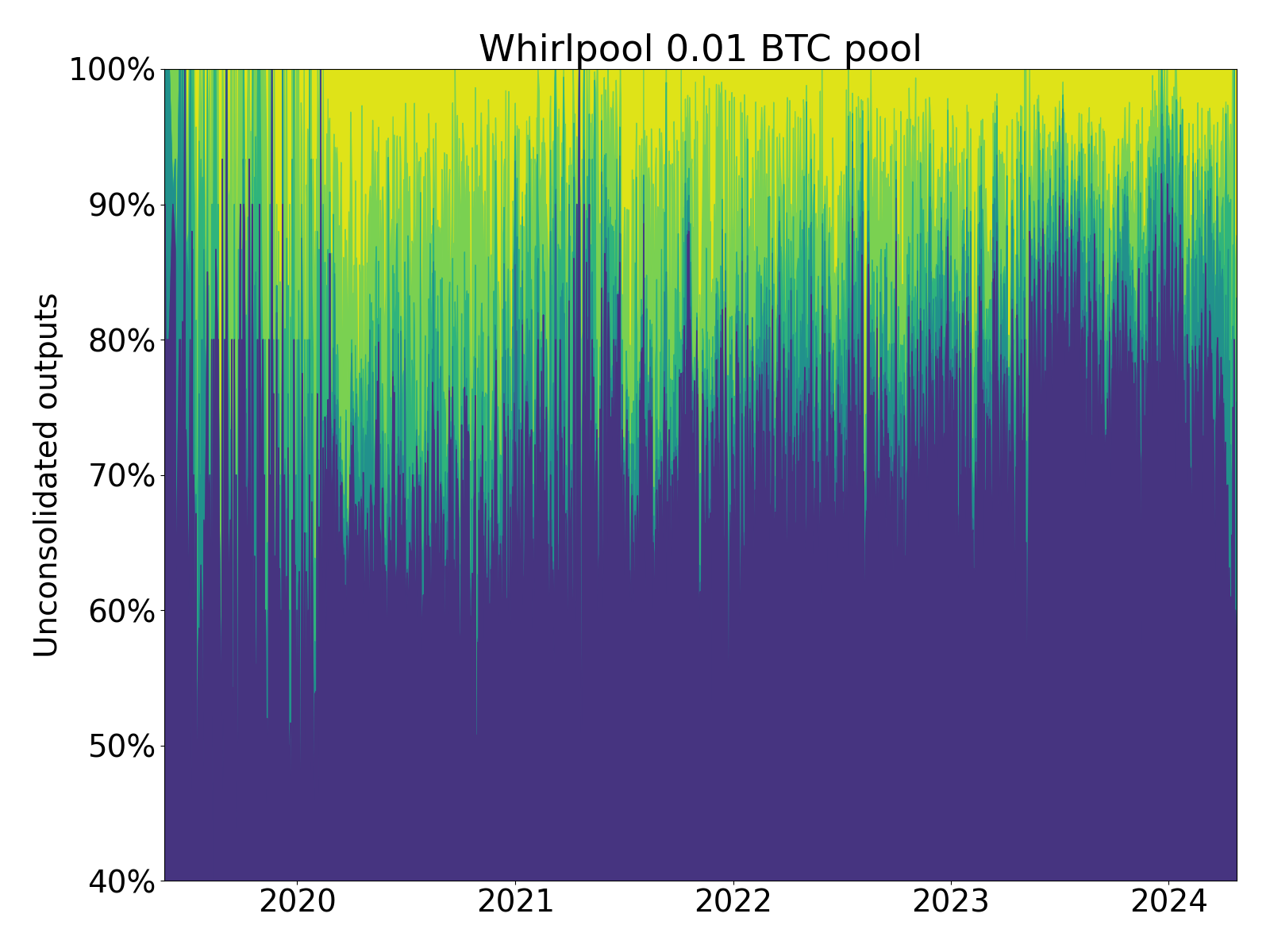}
        \put(14,8){\color{white}\tiny 0.01 BTC pool}
        \end{overpic}
    \end{minipage}
    \begin{minipage}[b]{0.24\textwidth}
        \centering
        \begin{overpic}[width=\linewidth,height=60px]{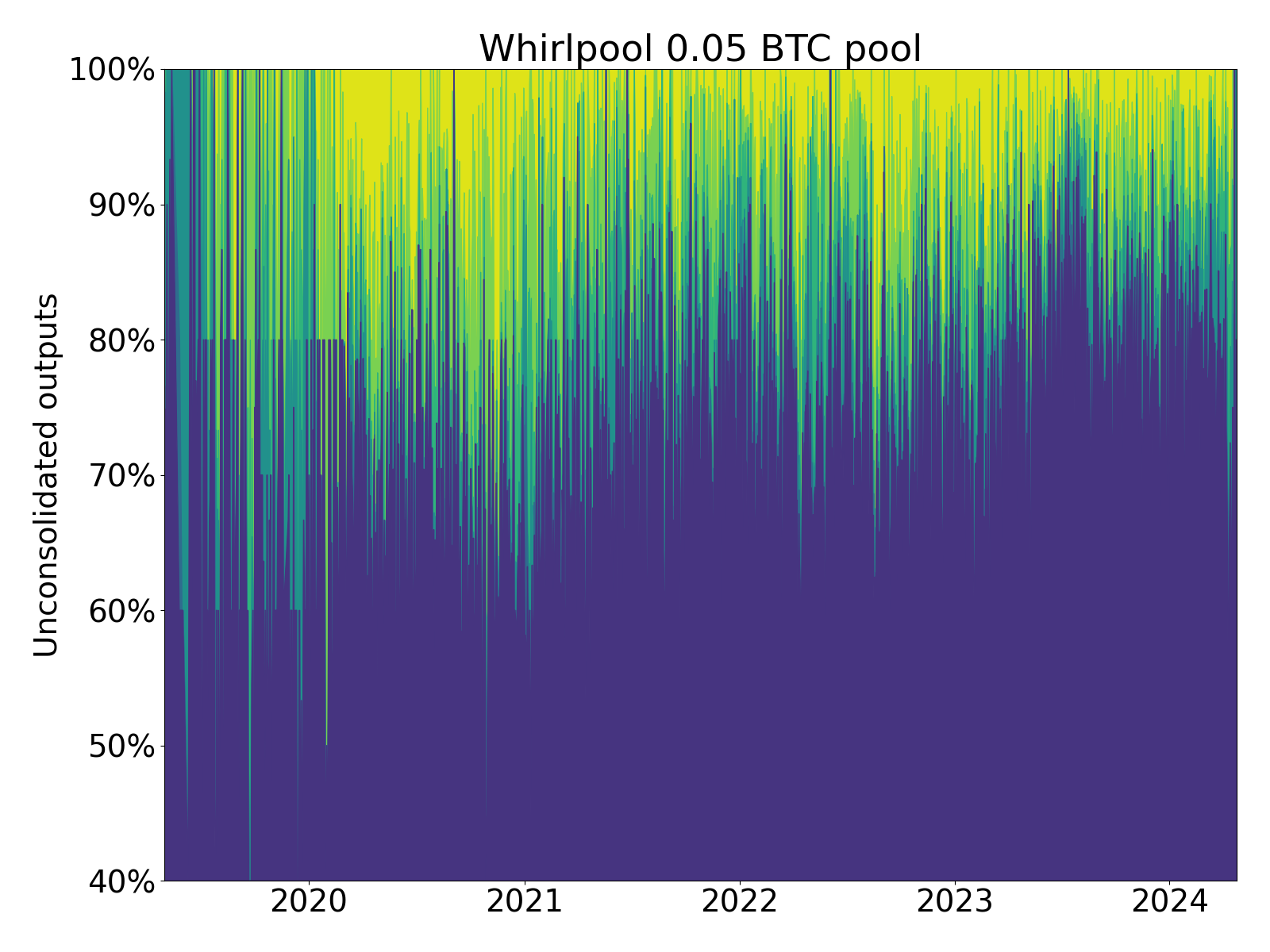}
        \put(14,8){\color{white}\tiny 0.05 BTC pool}
        \end{overpic}        
    \end{minipage}
\hfill
    \begin{minipage}[b]{0.24\textwidth}    
        \centering
        \begin{overpic}[width=\linewidth,height=60px]{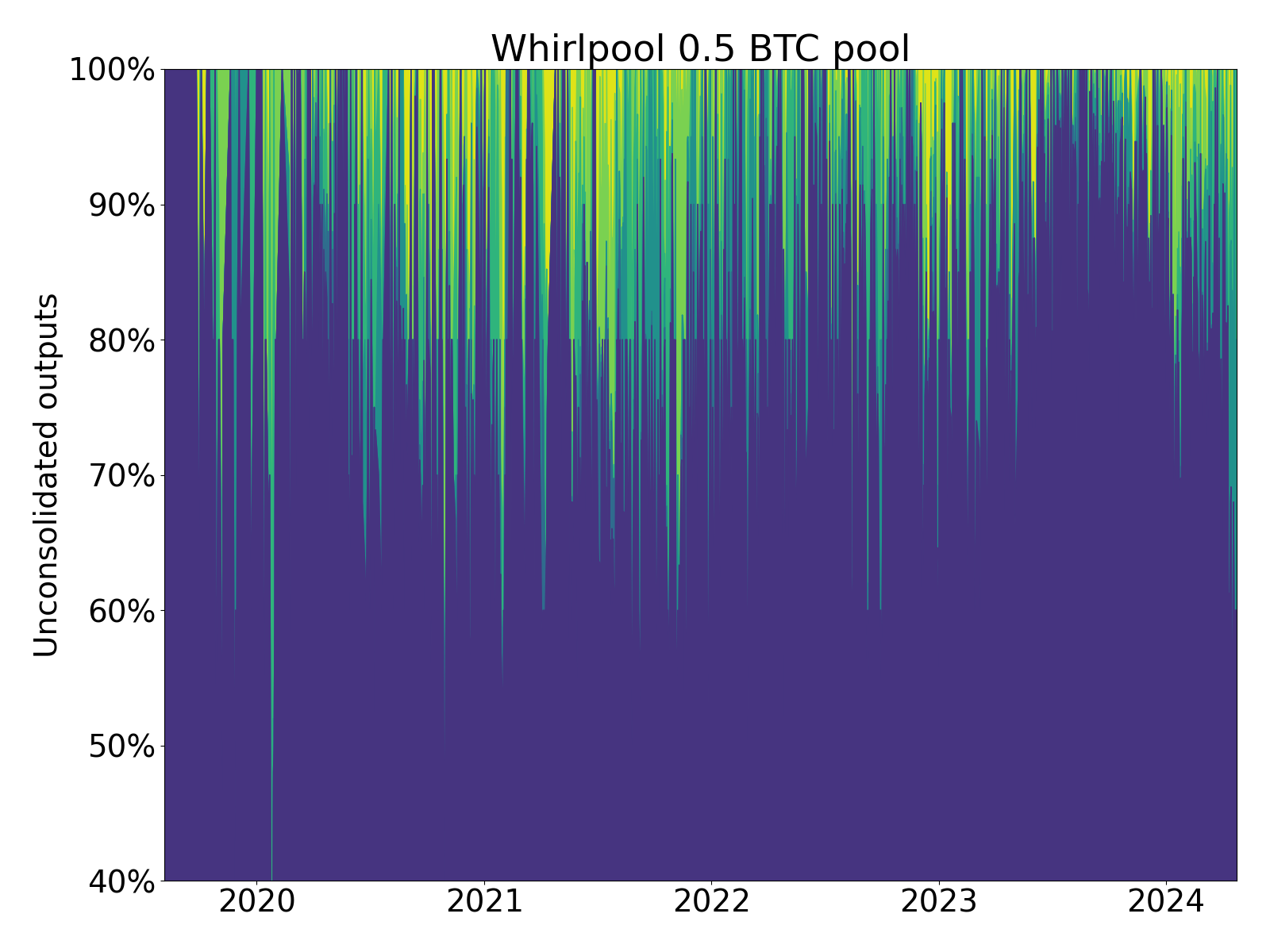}
        \put(14,8){\color{white}\tiny 0.5 BTC pool}
        \end{overpic}
    \end{minipage}

    \begin{minipage}[b]{0.24\textwidth}
    \centering
        \begin{overpic}[width=\linewidth,height=60px]{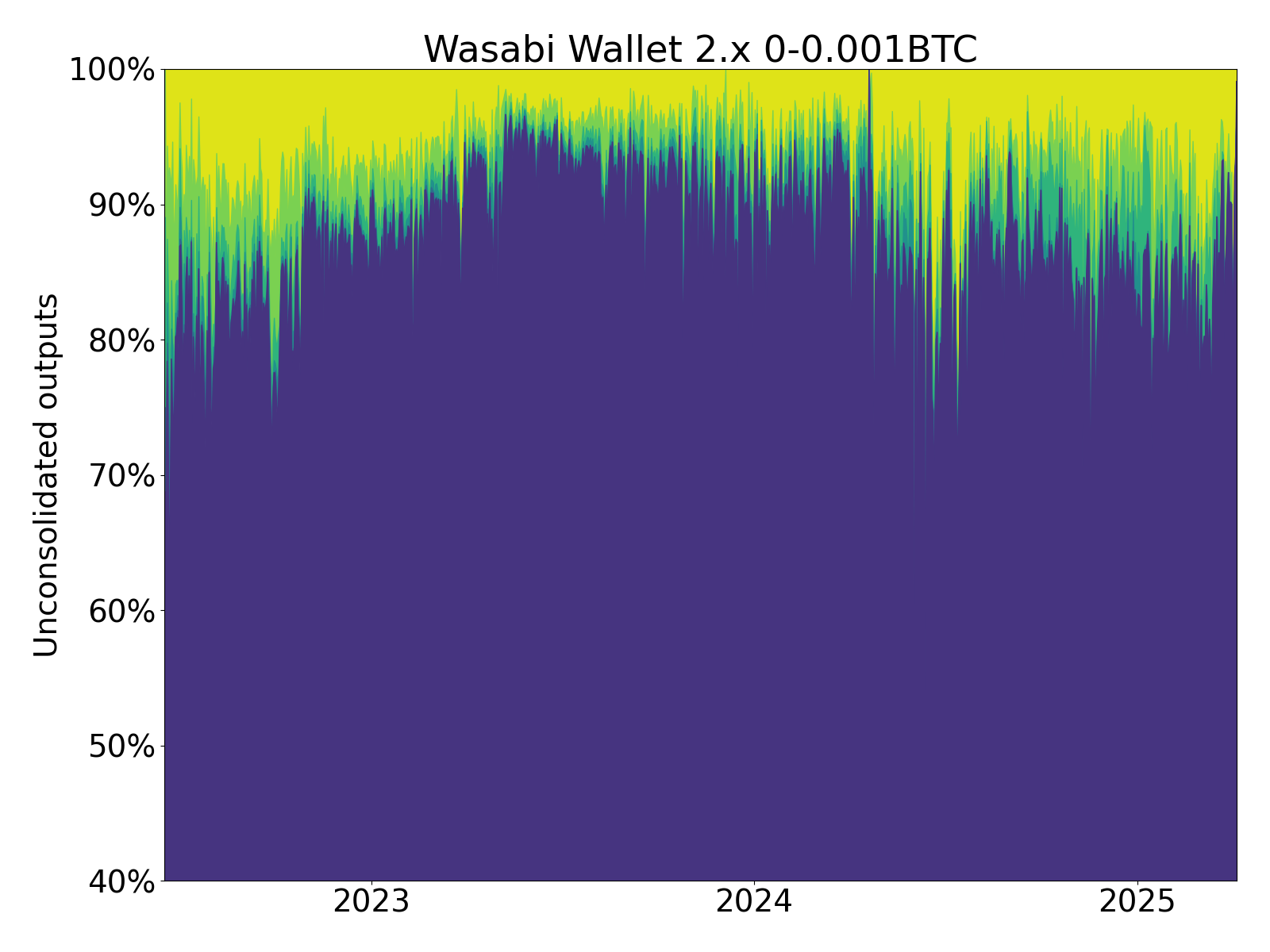}
        \put(14,9){\color{white}\tiny \( [0, 0.001] \) BTC}
         \put(-7,10){\rotatebox{90}{Wasabi 2.x}}
        \end{overpic}
    \end{minipage}
\hfill
    \begin{minipage}[b]{0.24\textwidth}
        \centering
        \begin{overpic}[width=\linewidth,height=60px]{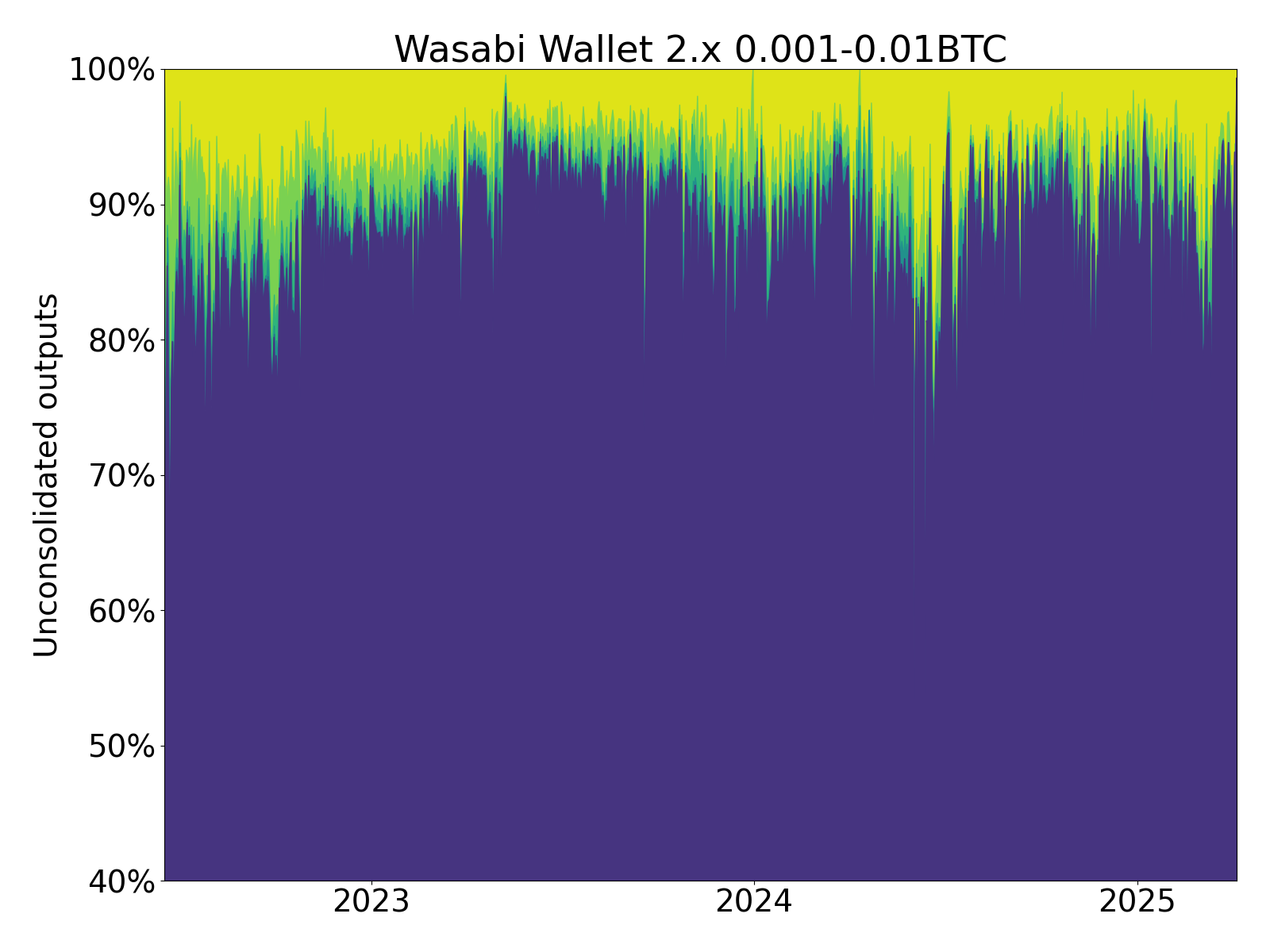}
        \put(14,9){\color{white}\tiny \( (0.001, 0.01] \) BTC}
        \end{overpic}
    \end{minipage}
    \begin{minipage}[b]{0.24\textwidth}
        \centering
        \begin{overpic}[width=\linewidth,height=60px]{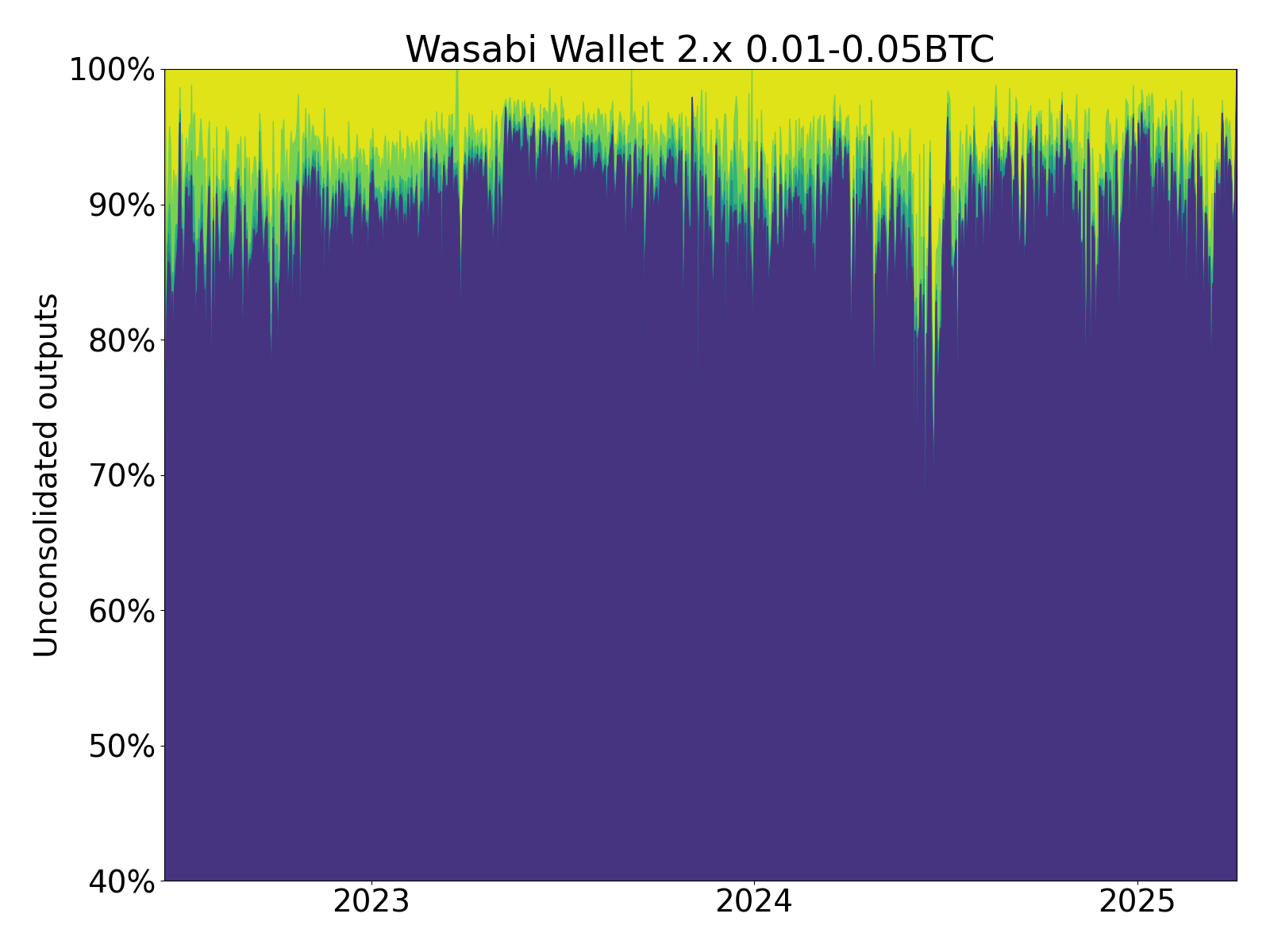}
        \put(14,9){\color{white}\tiny \( (0.01, 0.05] \) BTC}
        \end{overpic}        
    \end{minipage}
\hfill
    \begin{minipage}[b]{0.24\textwidth}    
        \centering
        \begin{overpic}[width=\linewidth,height=60px]{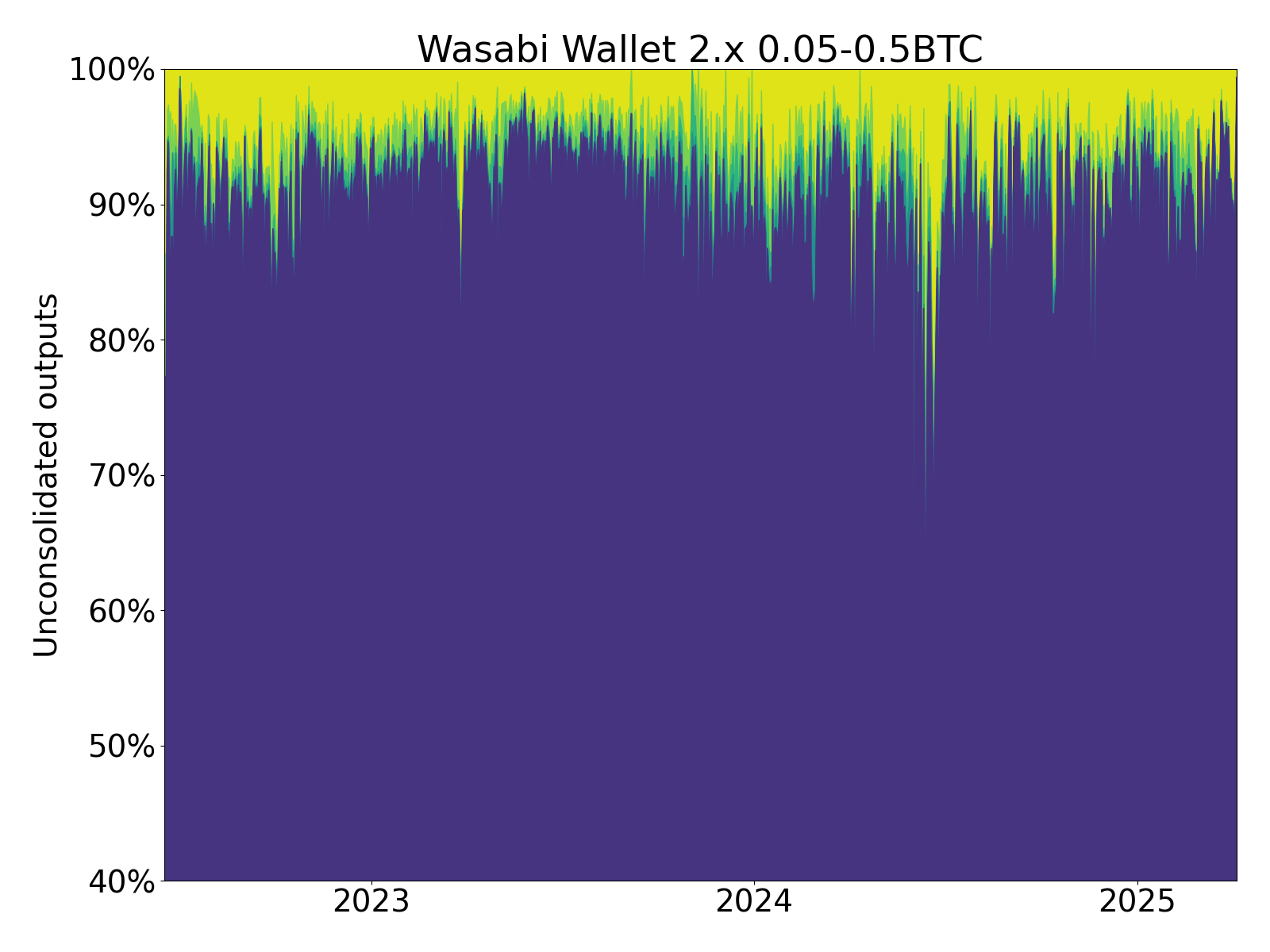}
        \put(14,9){\color{white}\tiny \( (0.05, 0.5] \) BTC }
        \end{overpic}
    \end{minipage}    
    \caption{Anonymity set loss for Wasabi 1.x, Whirlpool and Wasabi 2.x for different output denominations. All graphs have a y-axis range from 40\% to 100\%.}
    \label{fig:swdegradation}
\end{figure}

The anonymity set loss is not the same for different output denominations, as shown for all three investigated coinjoin designs in \Cref{fig:swdegradation}. The general trend is that smaller denominations are consolidated more frequently, resulting in higher anonymity set loss. Such observed behavior is logical -- the larger denominations have a higher chance of being alone enough for a subsequent payment, while smaller ones require joining multiple outputs to pay the desired amount.

 \section{Wasabi 2.x mappings enumeration}
 \label{app:WW2_mappings}
 \subsubsection{Wasabi 2.x Fees}
Wasabi 2.x has 3 types of fees -- mining fees, coordination fees, and fees paid because of imprecise decomposition of inputs into outputs (paid to miners or the coordinator). The coordination fees of 0.3\% can be easily subtracted from the relevant inputs. If we have a precise mining fee rate for a given transaction, we can also easily remove the mining fees. The decomposition fees cannot be removed as they are unpredictable. Therefore, we need to allow the sub-mappings to have the sum of inputs higher than the sum of outputs by this amount, plus some small margin to cover a possible error in mining fee computation. The upper limit on the decomposition fee is the minimal registrable output amount defined by the coordinator or the minimal value that is economical for the user, if it is higher.

\subsubsection{Wasabi 2.x mapping restrictions} 

If we assume that all users follow the protocol as implemented in Wasabi 2.x, then there are some limitations on the potential mappings of the resulting coinjoins. By analyzing the implementation of Wasabi 2.x, we identified the following restrictions on potential mappings.

Each user can register at most 10 inputs and 10 outputs. BTCPay~\cite{btcpay} has a coinjoin implementation compatible with Wasabi 2.x, which allows users to register up to 30 inputs. Therefore, we consider a limit of 30 inputs per user. 

Another factor limiting possible sub-mappings is that each wallet registers at most one change output, i.e., output that is not among the common denominations for a given coinjoin round. However, this restriction was recently loosened by the introduction of payments in coinjoins. As one user can have up to 4 payments in one coinjoin, there can be up to 5 outputs with values different from common denominations. Since we analyze only very small transactions, we decided not to implement this mapping restriction. 

\section{Discovered Wasabi 2.x implementation bugs}
\label{app:ww2_bugs}
When using Sake, we discovered that in some cases, the simulator never reproduces the same outputs for a wallet as observed in the emulation. We discovered that the simulator computes the common output denominations for all wallets at the same time, as it is assumed that all wallets should come up to the same values. However, the computation in Wasabi 2.x can sometimes produce different denominations for different wallets due to incorrect handling of coordination fees. This bug can lead to the linking of inputs to outputs if an attacker can find which combination of inputs could trigger such an incorrect computation of output denominations and observe such denominations in the outputs. 

This bug was fixed by removing coordination fees, and it is no longer affecting new coinjoins. However, it led us to discover a similar bug in a newer version of Wasabi 2.x, which occurred when a payment was used in the coinjoin. We reported this bug to the developers and helped to fix it.\footnote{\url{https://github.com/WalletWasabi/WalletWasabi/pull/13598}}

\section{Multiple coinjoin mappings enumeration}
\label{app:multiple_cj}
In practice, it is common to participate in multiple consecutive coinjoins. Our approach to mapping enumeration can be extended to analyze multiple interconnected transactions at once.
To analyze a set of transactions $\mathcal{T}$, we can create a new artificial transaction for which the mappings will then be enumerated. The artificial transaction will inherit all inputs/outputs of transactions that are not coming directly from/to another transaction in $\mathcal{T}$. 
For each pair of transactions from $\mathcal{T}$, we also need to know their order and the capacity of their link, i.e., the sum of outputs going from the first transaction to the second one. 

The enumeration then proceeds as usual, just filtering out the (sub-)mappings where the transfer of value between a pair of transactions would be higher than the capacity of their link.

\section{Mappings of real coinjoins}
\label{app:real_cjs}
We chose 5 small Wasabi 2.x coinjoins produced in 2024 by one of the currently running coordinators to test if the enumeration is practically applicable to real transactions. \Cref{tab:real_mappings} shows the number of numeric mappings for these transactions. The coordination fee charged by the coordinator was 0\%, and we used a range of possible mining fee rates as the mining fee rate used by wallets when choosing the output values can be different from the final fee rate of the whole transaction. However, it is possible to obtain the precise mining fee rate used in the transaction by regularly querying the coordinator's public API and logging this information. Note that the reported runtime of the mappings enumeration algorithm varies significantly (observed also for emulated coinjoins in \Cref{sect:mappings_enum}) and is not directly proportional to the coinjoin size, nor the number of mappings found. 
\begin{table}
    \centering
    \caption{Number of numeric mappings and execution time (T) in seconds and peak memory usage (PMU) in MBs for real Wasabi 2.x coinjoins by the \emph{opencoordinator.org} coordinator.}
    \begin{tabular}{|c|c|c|c|}
        \hline
        Transaction ID & \#Mappings & T & PMU\\ \hline
        {\fontsize{7pt}{7pt}\selectfont\href{https://mempool.space/tx/7e875be692881180ed3f322831615c280daf077bfd14bc120fb3c03dc6d381f6}{7e875be692881180ed3f322831615c280daf077bfd14bc120fb3c03dc6d381f6}} & 6565 & 35.5 & 33\\ \hline
        {\fontsize{7pt}{7pt}\selectfont\href{https://mempool.space/tx/f5f4fbf79355c777b9a83aef9202e85d4af83ffaf4b7f14e5ff6fd1b163eb3b0}{f5f4fbf79355c777b9a83aef9202e85d4af83ffaf4b7f14e5ff6fd1b163eb3b0}} & 38 & 0.07 & 36\\ \hline
        {\fontsize{7pt}{7pt}\selectfont\href{https://mempool.space/tx/c92a6046249fd613019e5dccb0f6c188e4d607eade2738250a10b7e1da2a0489}{c92a6046249fd613019e5dccb0f6c188e4d607eade2738250a10b7e1da2a0489}} & 2575 & 3.3 & 38 \\ \hline
        {\fontsize{7pt}{7pt}\selectfont\href{https://mempool.space/tx/a0ddfff8b16eaa9c461a15a2b70d174530b649426ac0472464b62f4a5ef02d6a}{a0ddfff8b16eaa9c461a15a2b70d174530b649426ac0472464b62f4a5ef02d6a}} & 34301 & 402 & 35 \\ \hline
        {\fontsize{7pt}{7pt}\selectfont\href{https://mempool.space/tx/ad5ad29922ab3067ecfcc608eba03c3d0aef08ce80a75e563ba8904875648743}{ad5ad29922ab3067ecfcc608eba03c3d0aef08ce80a75e563ba8904875648743}} & 17935 & 15998 & 42 \\ \hline

    \end{tabular}
    \label{tab:real_mappings}
\end{table}

\end{document}